\documentclass[aps,physrev,reprint,amsmath,amssymb,longbibliography,nofootinbib]{revtex4-2}
\usepackage[T1]{fontenc}
\usepackage[utf8]{inputenc}
\usepackage{lmodern}
\usepackage{amsmath,amssymb,amsthm,mathtools,booktabs}
\usepackage{graphicx}
\usepackage[colorlinks=true,linkcolor=blue,urlcolor=blue,citecolor=blue]{hyperref}
\hypersetup{pdftitle={Relax, Take Your Time, and Mind the Exponent: Algebraic Tails in Quantum Relaxation},pdfauthor={Marek Winczewski}}
\newtheorem{theorem}{Theorem}
\newtheorem{proposition}[theorem]{Proposition}
\newtheorem{corollary}[theorem]{Corollary}
\newtheorem{lemma}[theorem]{Lemma}
\newtheorem{observation}[theorem]{Observation}
\theoremstyle{remark}
\newtheorem{remark}[theorem]{Remark}
\newcommand{\tr}{\operatorname{tr}}
\newcommand{\Tr}{\operatorname{Tr}}
\newcommand{\id}{\mathrm{id}}
\newcommand{\E}{\mathcal E}
\newcommand{\D}{\mathcal D}
\newcommand{\spec}{\operatorname{spec}}
\newcommand{\norm}[1]{\left\lVert #1\right\rVert}

\newcommand{\dom}{\operatorname{Dom}}
\newcommand{\X}{\mathfrak S_1(\mathcal H)}
\AddToHook{env/thebibliography/begin}{\interlinepenalty=10000\relax}
\begin{document}
\title{Relax, Take Your Time, and Mind the Exponent:\\
Algebraic Tails in Quantum Relaxation}
\author{Marek Winczewski}
\email{Marek.Winczewski@ug.edu.pl}
\affiliation{Institute of Informatics, Faculty of Mathematics, Physics and Informatics,
University of Gda\'nsk, ul. Wita Stwosza 57, 80-308 Gda\'nsk, Poland}
\date{22 November 2026}
\begin{abstract}
We study how a damped oscillator responds to a reset that becomes
arbitrarily weak. For one-mode thermal attenuation, we show that the
time integral of the trace-norm error is finite exactly when the
initial state has a finite logarithmic particle-number moment, at every
fixed finite bath occupation and including states with number-basis
coherences. Applied to the reset state, the same criterion determines
whether the stationary response is differentiable at zero reset rate,
and whether the error at observation time $t$ is $O(t^{-1})$ for reset
rates $a_t/t$ with $a_t\to a>0$, uniformly over inputs of bounded mean
energy; in that regime the integrated operator response gives the
leading coefficient. A truncated logarithmic moment controls slower
decay: an explicit reset tail gives an error proportional to
$(\log t)/t$, and each fixed finite cutoff of that tail restores a
$1/t$ asymptote. A general forcing estimate also improves a pure-loss
strong-damping bound from $O((\log t)/t)$ to $O(t^{-1})$. A transverse
qubit provides a separate example, with an explicit $1/t$ coefficient
in its second-order cumulant dynamics.
\end{abstract}

\maketitle

\section{Introduction}
A weak reset can change the equilibrium of a dissipative system by
repeatedly populating levels that take a long time to empty. How fast
does that disturbance vanish as the reset rate tends to zero? At rate $r$, the stationary state is an
average of the relaxation trajectories that start from the reset state
$\sigma$. The typical time since the last reset is $1/r$. The
population of high levels can then control the stationary response
even when every initial state ultimately relaxes.

We study this balance through the family
\begin{equation}\label{eq:familyintro}
 \Phi_t=\exp(tL+D_t),
\end{equation}
where $L$ generates a relaxing quantum Markov semigroup and $D_t$ is
a bounded operator on the trace class. For each $t>0$, the exponential
evolves for a duration $t$ with the generator $L+D_t/t$ held fixed. If
$D_t$ is uniformly bounded, its contribution to the generator is
$O(t^{-1})$. In the reset examples, increasing $t$ both lengthens the
protocol and lowers the reset rate. For a fixed correction $D_t=D$,
the same exponential can be read at unit duration with the damping
generator scaled to $tL$.

For a thermally damped oscillator, the dividing line is the logarithmic
particle-number moment of the reset state. Its finiteness is equivalent
both to a finite stationary derivative at zero reset rate and to an
$O(t^{-1})$ error along the reset family, uniformly over inputs of
bounded mean energy. The moment determines when this inverse-time
expansion exists. The full integrated relaxation of the reset state
determines its coefficient and therefore retains information about
temperature and coherences.

The logarithm comes from a simple timescale. Under pure loss, the mean
time for the last of $n$ excitations to disappear is proportional to the
harmonic number $H_n\sim\log n$. Related moment conditions occur in
batch-arrival queues~\cite{Cong1994,Yajima2016}. The quantum problem is to
control coherences with this same moment. At positive temperature one
must also show that population deviations of opposite signs cannot
cancel a divergent response. A vacuum-population identity supplies
the needed lower bound.

The renewal description of resetting and its stationary state are
established~\cite{EvansMajumdar2011,EvansMajumdarSchehr2020,MukherjeeSenguptaMajumdar2018,RoseEtAl2018,PerfettoCarolloLesanovsky2022},
and perturbation theory for quantum reset models is developed in
Ref.~\cite{HaackJoye2021}. Here the logarithmic threshold identifies when
the usual finite-dimensional picture breaks down. A finite-dimensional
relaxing generator has an analytic trace-one stationary branch near
zero strength of a fixed trace-annihilating perturbation. More
generally, uniform exponential relaxation gives a bounded
reduced-inverse response~\cite{SzehrWolf2013,BenattiNagyNarnhofer2011}.
For the oscillator, arbitrarily slow relaxation across the input
state space makes the integrability of the particular reset trajectory
decisive.

The strong-damping viewpoint connects \eqref{eq:familyintro} to
singular semigroup perturbations~\cite{Kurtz1973},
dissipation-projected dynamics~\cite{ZanardiCamposVenuti2014,ZanardiCamposVenuti2015},
and strong-coupling limits~\cite{BurgarthEtAl2019,Salzmann2024}.
The corresponding discrete limits arise in quantum Zeno
products~\cite{BeckerDattaSalzmann2021,BurgarthEtAl2020,MobusRouze2023,Mobus2026}.
Our rate comparison concerns the continuous pure-loss setting:
for fixed Hermiticity-preserving, trace-annihilating perturbations
under the finite-energy hypotheses of Ref.~\cite{Salzmann2024},
the strong-damping bound improves from $O((\log t)/t)$ to $O(t^{-1})$.
A logarithmic moment of the forcing already suffices
(Appendix~\ref{app:oscforcing}).

\begin{figure*}[t]
\centering
\includegraphics[width=\textwidth]{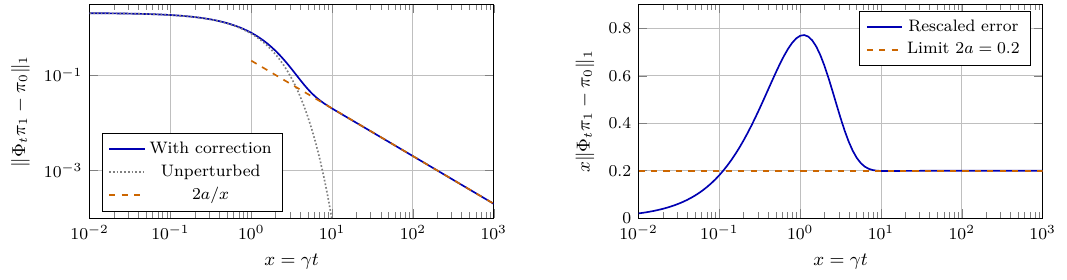}
\caption{An algebraic tail along the family $\Phi_t=e^{tL+D_t}$,
caused by the displacement of its frozen stationary state.
The initial state is $\pi_1$ and the reset strength is $a=0.1$, with
$a_t=a(1-e^{-\gamma t})$. Left: the exact error
\eqref{eq:excitederror}, the unperturbed decay $2e^{-\gamma t}$, and the
asymptote $2a/(\gamma t)$. Right: the rescaled error converges to the
nonzero coefficient $2a$. Both panels use the exact infinite-dimensional
solution, without a Fock-space truncation.}
\label{fig:inverse}
\end{figure*}

We start with an exact one-photon example and use it to isolate the
forcing that governs the oscillator estimates
(Sections~\ref{sec:example} and~\ref{sec:strong}).
Sections~\ref{sec:oscillator} and~\ref{sec:logtail} establish the
logarithmic threshold and the rates for heavier tails.
Section~\ref{sec:cumulant} returns to the same inverse-time law in a
qubit whose exponent comes from a system--bath interaction. The
appendices collect the proofs and additional examples.

\section{How an inverse-time tail arises}\label{sec:example}
A single excitation shows how exponentially fast damping can leave an
inverse-time error in a reset protocol. Consider a harmonic oscillator
with loss rate $\gamma>0$ and phase rotation of frequency $\omega$.
We measure errors in the trace norm $\norm{\cdot}_1$, twice the
conventional trace distance.
Write $\hat a$ for the annihilation operator, $N=\hat a^\dagger\hat a$,
and $\pi_n=|n\rangle\langle n|$. The semigroup $T_s$ is the pure-loss
attenuator of transmissivity $e^{-\gamma s}$ followed by this rotation.
Its generator is $L=-i\omega[N,\,\cdot\,]+\gamma\D[\hat a]$, with
the dissipator convention
\begin{equation}\label{eq:dissipator}
 \D[V](X)=VXV^\dagger-\tfrac12\{V^\dagger V,X\}.
\end{equation}
The vacuum is invariant. With one photon initially present, the survival
probability is $e^{-\gamma s}$, so
$T_s\pi_1=\pi_0+e^{-\gamma s}(\pi_1-\pi_0)$.

For a state $\sigma$, let $P_\sigma(X)=\tr(X)\sigma$ be the replacement
channel. We add a reset toward $\pi_1$ by choosing
\begin{equation}\label{eq:switchon}
 a_t=a(1-e^{-\gamma t}),\qquad
 D_t=a_t(P_{\pi_1}-\id),\qquad a>0.
\end{equation}
The perturbation obeys $\norm{D_t}\le2a$. Each frozen generator
$L+D_t/t$ generates a completely positive, trace-preserving (CPTP)
semigroup, and the continuous switch-on
makes $\Phi_0=\id$. For the one-photon input, the
exact solution gives
\begin{align}
 p_t&=\frac{a_t}{\gamma t+a_t}(1-e^{-\gamma t-a_t}),
 \label{eq:onephotonp}\\
 \norm{\Phi_t\pi_1-\pi_0}_1
 &=2e^{-\gamma t-a_t}+2p_t.
 \label{eq:excitederror}
\end{align}
Here $p_t$ is the population supplied by the reset; for vacuum input,
the output is $(1-p_t)\pi_0+p_t\pi_1$.

The first term in \eqref{eq:excitederror} comes from the initial photon.
The second comes from resetting. In the frozen protocol, pumping at
rate $a_t/t$ and loss at rate $\gamma$ select a stationary excited
population $a_t/(\gamma t+a_t)$. The protocol approaches that value
exponentially, and the value itself falls as $a/(\gamma t)$ as the
protocol is lengthened. Equivalently, a reset within the last damping
time has probability of order $a/(\gamma t)$ and can leave a photon
at the output.

The same coefficient holds for every initial state of finite mean
energy:
\begin{equation}\label{eq:onephotonsimple}
 \norm{\Phi_t\rho-\pi_0}_1\sim\frac{2a}{\gamma t}.
\end{equation}
This asymptotic is uniform when $\tr(N\rho)$ is bounded.
Appendix~\ref{app:exactinverse} gives the exact map and the uniform
limits. Figure~\ref{fig:inverse} illustrates how even a small bounded
correction eventually determines the leading error. We next ask which
features of this example persist for a general relaxing dynamics.

\section{The forcing that controls relaxation}\label{sec:strong}
The one-photon example suggests looking at what the perturbation
does to the stationary state. To make that idea precise, let $T_s$,
$s\ge0$, be a strongly continuous CPTP semigroup on the trace class
$\X$ of a separable Hilbert space, with generator $L$. The maps
are complex-linear and act in the Schr\"odinger picture. We assume
there is a density operator $\rho_*$ such that, for every initial
state $\rho$,
\begin{equation}\label{eq:relax}
 \norm{T_s\rho-\rho_*}_1\longrightarrow0\qquad(s\to\infty).
\end{equation}
Then $\rho_*$ is invariant and $T_sX\to PX$ for every trace-class $X$,
where $P(X)=\tr(X)\rho_*$. Criteria for this type of relaxation
are given in Refs.~\cite{Spohn1977,Frigerio1982,Fagnola2008}.
Vector norms below are trace norms and map norms are the corresponding
induced norms on the complex trace class.

For each $t>0$, take a bounded operator $D_t$ satisfying
\begin{equation}\label{eq:structure}
 D_t(X^\dagger)=D_t(X)^\dagger,\qquad \tr D_t(X)=0
 \quad(X\in\X).
\end{equation}
The bounded perturbation theorem defines $\Phi_t$ as the time-$t$
value of the semigroup generated by $L+D_t/t$, on $\dom L$
\cite{EngelNagel2000}. This construction preserves Hermiticity and trace.
Positivity of the output requires a separate argument, as in the reset
example above. Background and domain details are
collected in Appendix~\ref{app:framework}; see also Ref.~\cite{Wolf2012}.

The initial deviation $\rho-\rho_*$ contributes a transient. Even when
that deviation is zero, the perturbation creates a displacement through
its action $b_t=D_t\rho_*$ on the stationary state. The estimate below
uses variation of constants to bound the contribution of this forcing
by its accumulated relaxation under $T_s$.

\begin{theorem}[Stability through the forcing]\label{thm:strong}
Assume \eqref{eq:relax} and \eqref{eq:structure}, with
$\sup_{t\ge t_0}\norm{D_t}\le K<\infty$. For a state $\rho$, put
$x=\rho-\rho_*$ and $b_t=D_t\rho_*$. Then, for $t\ge t_0$,
\begin{align}\label{eq:cesaroestimate}
 \norm{\Phi_t\rho-\rho_*}_1
 &\le\norm{T_tx}_1
 +\frac{Ke^K}{t}\int_0^t\norm{T_sx}_1\,ds\notag\\
 &\hspace{9mm}+\frac{e^K}{t}\int_0^t\norm{T_sb_t}_1\,ds.
\end{align}
Consequently, $\Phi_t\rho\to\rho_*$ for every initial state if
\begin{equation}\label{eq:forcingcesaro}
 \frac1t\int_0^t\norm{T_sD_t\rho_*}_1\,ds\longrightarrow0.
\end{equation}
Relative compactness of $\{D_t\rho_*:t\ge t_0\}$ in trace norm is a
sufficient condition. In particular, every fixed bounded $D_t=D$
satisfying \eqref{eq:structure} preserves strong relaxation.
\end{theorem}

For each fixed input, the first two terms in \eqref{eq:cesaroestimate}
vanish by strong relaxation and Ces\`aro averaging; a rate for them
requires control of the initial-state relaxation. The last term isolates
the forcing: if $\int_0^t\norm{T_sb_t}_1ds$ stays bounded, its
contribution is $O(t^{-1})$. Compactness is needed only for the forcing
set, not for the operators $D_t$. The proof and an energy-based
compactness criterion are given in Appendices~\ref{app:strong} and
\ref{app:compact}. For fixed $D$, convergence also follows from the
general strong-damping theorem of Ref.~\cite{Salzmann2024}.

\subsection{The uniform-relaxation benchmark}\label{sec:uniform}
The forcing has the usual linear-response interpretation when
$A=L|_{\E}$, $\E=\ker\tr$, satisfies
\begin{equation}\label{eq:basebound}
 \norm{e^{sA}}\le M e^{-\gamma s},\qquad M\ge1,\quad\gamma>0.
\end{equation}
Then $A^{-1}=-\int_0^\infty e^{sA}ds$ is bounded, and for uniformly
bounded $D_t$ standard perturbation estimates give, uniformly over inputs,
\begin{equation}\label{eq:leading}
 \begin{split}
 \Phi_t\rho-\rho_*={}&-\frac1tA^{-1}D_t\rho_*\\
 &+O(t^{-2})+O(e^{-\gamma t}).
 \end{split}
\end{equation}
This is the reduced-inverse response
\cite{BenattiNagyNarnhofer2011,SzehrWolf2013}. The exact displacement,
sublinear-perturbation version, and quantitative bounds are collected
in Appendices~\ref{app:uniform} and \ref{app:response}; spectral and
positivity consequences are in Appendices~\ref{app:finite} and
\ref{app:positivity}. A relaxing finite-dimensional semigroup satisfies
\eqref{eq:basebound}. The oscillator does not: its response must instead
be controlled on the particular forcing supplied by the perturbation.

\section{The oscillator and the logarithmic threshold}\label{sec:oscillator}
We extend the pure-loss model to a bath with fixed finite mean occupation
$0\le\bar n<\infty$. The thermal attenuation semigroup has the formal
generator
\begin{equation}\label{eq:oscillatorL}
 L=-i\omega[N,\,\cdot\,]
 +\gamma(\bar n+1)\D[\hat a]+\gamma\bar n\D[\hat a^\dagger].
\end{equation}
Here $L$ is the generator of the conservative semigroup on the trace
class; the displayed expression is well defined on finite linear
combinations of number-state matrix units. Equivalently, $T_s$ is a
thermal attenuator of transmissivity $e^{-\gamma s}$ followed by the
phase rotation. Its invariant state is
\begin{equation}\label{eq:thermalstate}
 \tau_{\bar n}=\sum_{j=0}^\infty
 \frac{\bar n^j}{(\bar n+1)^{j+1}}\pi_j,
 \qquad\tau_0=\pi_0.
\end{equation}
We need to bound the error accumulated over the whole relaxation
trajectory of the reset state. Our estimates use the beam-splitter
representation of attenuation~\cite{LamiPlenioGiovannettiHolevo2020}.

\subsection{Why a logarithmic moment appears}
For a nonnegative self-adjoint observable $V$, we use the extended
expectation $\tr(V\rho)=\sup_{m>0}\tr[\min(V,mI)\rho]$.
This convention also covers infinite moments without requiring
$V\rho$ itself to be trace class. For $0\le E<\infty$, write
$\mathcal S_E=\{\rho\ge0:\tr\rho=1,\ \tr(N\rho)\le E\}$.
Finite mean energy provides a uniform transient estimate:
\begin{equation}\label{eq:energyshort}
 \sup_{\rho\in\mathcal S_E}\norm{T_s\rho-\tau_{\bar n}}_1
 \le C_Ee^{-\gamma s/2},
\end{equation}
where $C_E=\pi\sqrt{(2\bar n+1)E+\bar n}$. Nevertheless,
\begin{equation}\label{eq:oscnotuniform}
 \sup_{\rho\text{ a state}}\norm{T_s\rho-\tau_{\bar n}}_1=2
 \qquad(0\le s<\infty).
\end{equation}
Arbitrarily energetic inputs retain a distinguishable excitation at any
fixed time. Each fixed state still relaxes, by approximation with finite
energy states. The exponent $\gamma/2$ cannot be improved in general:
at zero temperature a nonzero coherent input has distance
$2\sqrt{1-\exp(-|\alpha|^2e^{-\gamma s})}
\sim2|\alpha|e^{-\gamma s/2}$, as proved in Appendix~\ref{app:oscenergy}.
Bertini, De Sole, and Posta obtain quantitative exponential relaxation
for bosonic semigroups, including coherences, on classes with factorial
bounds on the joint particle-number
moments~\cite[Eq.~(5.9) and Theorem~5.4]{BertiniDeSolePosta2024}.
The time-integrated error considered here requires much weaker control.

For pure loss, define
$h_{\pi_n}=\int_0^\infty(T_s\pi_n-\pi_0)\,ds$. Then
\begin{equation}\label{eq:harmonicresponse}
 \norm{h_{\pi_n}}_1=\frac{2H_n}{\gamma},\qquad
 H_n=\sum_{j=1}^n\frac1j,\quad H_0=0.
\end{equation}
The time for all $n$ independently decaying excitations to disappear has
mean $H_n/\gamma$. The logarithmic growth $H_n\sim\log n$ therefore
suggests the relevant moment before any trace-norm estimate is made.
The identity and its extension to diagonal resets are derived in
Appendices~\ref{app:energyresponse} and \ref{app:logmoment}.

For an arbitrary density operator $\sigma$, including coherences, put
\begin{equation}\label{eq:logmoment}
 H_N=\sum_{n\ge0}H_n\pi_n,\qquad
 \mathcal M_\sigma=\tr(H_N\sigma).
\end{equation}
We allow $\mathcal M_\sigma=\infty$.
Since $\log(n+1)\le H_n\le1+\log n$ for $n\ge1$, this moment is finite
exactly when $\tr[\sigma\log(I+N)]<\infty$.

\subsection{An exact criterion for reset perturbations}\label{sec:thermalmoment}
Fix a reset state $\sigma$ and take $D_t=a_t(P_\sigma-\id)$, where
$0\le a_t\le A_0<\infty$ and $a_t\to a>0$. Each $\Phi_t$ is CPTP.
The standard renewal representation
\cite{MukherjeeSenguptaMajumdar2018,RoseEtAl2018,PerfettoCarolloLesanovsky2022}
separates the initial transient from the reset contribution:
\begin{equation}\label{eq:resetresponse}
\begin{split}
 \Phi_t\rho-\tau_{\bar n}
 ={}&e^{-a_t}(T_t\rho-\tau_{\bar n})\\
 &+\frac{a_t}{t}\int_0^t e^{-a_ts/t}(T_s\sigma-\tau_{\bar n})\,ds.
\end{split}
\end{equation}
The two states play different roles: the energy bound on the input
$\rho$ controls the first term uniformly, whereas the tail of the reset
state $\sigma$ decides whether the second term has a finite
inverse-time coefficient. To express this threshold, define
\begin{equation}\label{eq:thermalI}
 I_{\bar n}(\sigma)=\int_0^\infty
        \norm{T_s\sigma-\tau_{\bar n}}_1\,ds.
\end{equation}

\begin{theorem}[The logarithmic-moment threshold]\label{thm:thermalmoment}
For the unperturbed thermal attenuation semigroup, any fixed
$0\le\bar n<\infty$, and any density operator $\sigma$,
\begin{equation}\label{eq:thermaliff}
 I_{\bar n}(\sigma)<\infty
 \quad\Longleftrightarrow\quad\mathcal M_\sigma<\infty.
\end{equation}
For the reset family just defined and every $0\le E<\infty$,
$\Phi_t\rho\to\tau_{\bar n}$ uniformly on $\mathcal S_E$, and
\begin{equation}\label{eq:thermalresetiff}
 \sup_{\rho\in\mathcal S_E}\norm{\Phi_t\rho-\tau_{\bar n}}_1=O(t^{-1})
 \quad\Longleftrightarrow\quad\mathcal M_\sigma<\infty.
\end{equation}
When the moment is finite, the trace-class response
\begin{equation}\label{eq:thermalresponse}
 h_\sigma^{(\bar n)}=\int_0^\infty(T_s\sigma-\tau_{\bar n})\,ds
\end{equation}
belongs to $\dom L$, satisfies
$Lh_\sigma^{(\bar n)}=\tau_{\bar n}-\sigma$, and obeys
\begin{equation}\label{eq:thermalresetlimit}
 \sup_{\rho\in\mathcal S_E}
 \norm{t(\Phi_t\rho-\tau_{\bar n})-a h_\sigma^{(\bar n)}}_1
 \longrightarrow0.
\end{equation}
Thus the trace-norm coefficient is $a\norm{h_\sigma^{(\bar n)}}_1$,
which is strictly positive exactly when $\sigma\ne\tau_{\bar n}$.
\end{theorem}

The threshold involves only the number populations of the reset state.
The coefficient also depends on the coherences and on temperature:
the integral of the norm in \eqref{eq:thermalI} controls
integrability, and the norm of the operator integral in
\eqref{eq:thermalresponse} sets the coefficient. Equation
\eqref{eq:thermalresetlimit} gives an $o(t^{-1})$ error after subtracting
$a h_\sigma^{(\bar n)}/t$ from the unscaled state displacement,
uniformly on $\mathcal S_E$.

At positive temperature the vacuum is no longer absorbing, and its
population can approach equilibrium from either side. The key identity is
\begin{equation}\label{eq:thermalvacidentity}
 \int_0^\infty\tr[\pi_0(\tau_{\bar n}-T_s\sigma)]\,ds
 =\frac{\mathcal M_\sigma-\log(1+\bar n)}{\gamma(1+\bar n)}.
\end{equation}
The negative part of the integrand is integrable, so this identity also
holds with value $+\infty$. An infinite logarithmic moment therefore
cannot be hidden by cancellations in the reset average. This proves
necessity; sufficiency uses the standard decomposition into pure loss
followed by a quantum-limited amplifier~\cite{GarciaPatronEtAl2012}.
In fact, when $\mathcal M_\sigma=\infty$,
$\inf_\rho t\norm{\Phi_t\rho-\tau_{\bar n}}_1\to\infty$,
where the infimum is over all input states. The proof and quantitative
bounds are in Appendix~\ref{app:thermalmoment}. Moment-based
perturbation theory for bosonic evolutions~\cite{GondolfMobusRouze2024}
provides a related framework; here the reset target may be non-Gaussian
and need only have a logarithmic moment.

Three further cases are worked out in the appendices: an
infinite-energy reset can still satisfy the criterion
(Appendix~\ref{app:logmoment}), resetting to the vacuum in a thermal
bath gives a nonzero $1/t$ coefficient (Appendix~\ref{app:thermalvacuum}),
and coherences change that coefficient even at fixed number populations
(Appendix~\ref{app:coherentreset}).

\subsection{The stationary response to weak resetting}\label{sec:weakreset}
The same threshold governs the stationary state at weak reset rates.
At any fixed $r>0$, the renewal formula gives the unique stationary
state of $L+r(P_\sigma-\id)$
\cite{EvansMajumdarSchehr2020,RoseEtAl2018},
\begin{equation}\label{eq:resetness}
 \rho_r^{\mathrm{ss}}=r\int_0^\infty e^{-rs}T_s\sigma\,ds
                    =r(r-L)^{-1}\sigma.
\end{equation}
As $r\downarrow0$, this state has the linear response
\begin{equation}\label{eq:nessresponse}
 \frac{\rho_r^{\mathrm{ss}}-\tau_{\bar n}}r
 \longrightarrow h_\sigma^{(\bar n)}
 \quad\text{if }\mathcal M_\sigma<\infty.
\end{equation}
If the moment is infinite, the norm of the left-hand side diverges.
Thus the criterion is exactly the condition for a finite trace-norm
linear response to the reset rate.

For $a_t>0$, put $r=a_t/t$. The exact decomposition is
\begin{equation}\label{eq:nessdecomposition}
 \Phi_t\rho=\rho_r^{\mathrm{ss}}
       +e^{-a_t}T_t(\rho-\rho_r^{\mathrm{ss}}).
\end{equation}
With a finite logarithmic moment, the last term is $o(t^{-1})$
uniformly on $\mathcal S_E$. This identifies the inverse-time
coefficient with the stationary susceptibility. For heavier tails,
that conclusion requires care: the last term may be of the same order
as the stationary displacement. Appendix~\ref{app:weakreset} proves
these statements and gives an explicit class with different leading
coefficients for the finite observation time and the stationary limit.

\subsection{Beyond a fixed reset state}
The integrated-response argument also applies to general bounded
perturbations. Under \eqref{eq:structure}, with
$\sup_{t\ge t_0}\norm{D_t}\le K$, a sufficient condition is
\begin{equation}\label{eq:oscforcingmoment}
 F_H=\sup_{t\ge t_0}\tr\bigl(H_N|D_t\tau_{\bar n}|\bigr)<\infty.
\end{equation}
For each finite $E$, it gives
\begin{equation}\label{eq:generalforcingrate}
 \sup_{\rho\in\mathcal S_E}\norm{\Phi_t\rho-\tau_{\bar n}}_1
 \le C_Ee^{-\gamma t/2}+\frac{C}{t},
\end{equation}
with $C$ independent of $t$ and $\rho$. The explicit constant is in
Appendix~\ref{app:oscforcing}. Unlike the reset criterion,
\eqref{eq:oscforcingmoment} is only a sufficient condition for general
signed forcing. Appendix~\ref{app:oscforcing} also gives the precise
comparison with the attenuator bound discussed in the Introduction.

For moving reset states $\sigma_t$, a uniform mean-energy bound also
ensures $O(t^{-1})$ on $\mathcal S_E$ (Appendix~\ref{app:oscreset}).
Some uniform control of their tails is essential: targets whose excitation
number increases too quickly with $t$ can prevent relaxation altogether,
despite bounded perturbations and CPTP maps
(Appendix~\ref{app:increasing}). For a fixed reset state, convergence
always holds; what may fail is the inverse-time upper bound.

For a finite product of independent thermal attenuators and a product
reset target, a short tensor-product argument
(Appendix~\ref{app:multimode}) extends the integrability criterion to a
finite logarithmic moment of the total occupation $N_{\mathrm{tot}}$.

\begin{figure*}[t]
\centering
\includegraphics[width=\textwidth]{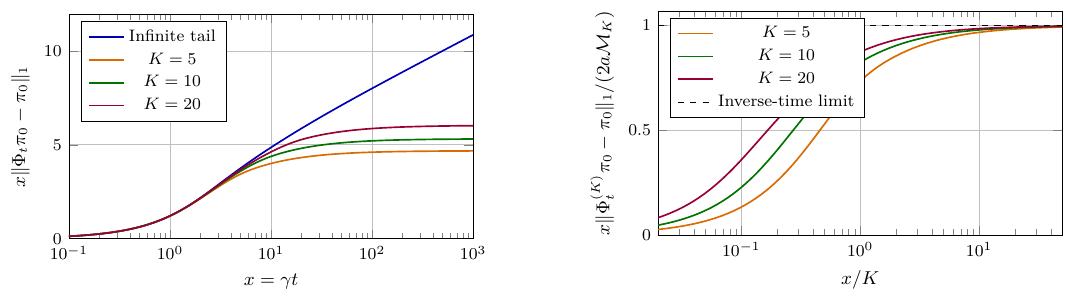}
\caption{A heavy reset tail and its finite-cutoff counterparts.
The reset strength is $a=1$, the input is the vacuum, and $x=\gamma t$.
The infinite reset has weights $p_k=6/(\pi^2k^2)$ on
$\pi_{\lceil e^k\rceil}$. Its normalized cutoff has
$p_k^{(K)}=k^{-2}/\sum_{j=1}^Kj^{-2}$ for $1\le k\le K$,
with $\mathcal M_K=\sum_{k=1}^Kp_k^{(K)}H_{\lceil e^k\rceil}$.
Left: the rescaled error continues to grow for the infinite tail,
whereas each fixed cutoff approaches $2a\mathcal M_K$, equal to approximately
$4.68474$, $5.32180$, and $6.03865$ for $K=5,10,20$. Right: dividing by this coefficient displays recovery of the
inverse-time law for each finite cutoff; the horizontal scale
compares evolution time with the logarithm of its largest occupied level.
Curves are numerical quadratures of the exact renewal formula.
The infinite sum is evaluated with a bounded tail remainder;
quadrature refinement changes the displayed quantities by less than $10^{-8}$.
No truncation of the oscillator generator is used.}
\label{fig:tailcutoff}
\end{figure*}

\section{From the injected tail to the relaxation rate}\label{sec:logtail}
The threshold separates finite and divergent integrated response, but
does not by itself specify the slower rate. For pure loss this rate can
be bounded directly in terms of the portion of the moment accessible
by time $t$. Keep $a>0$ fixed and $D=a(P_\sigma-\id)$, and define the
truncated harmonic moment
\begin{equation}\label{eq:truncatedmoment}
 m_\sigma(x)=\tr[\sigma\min(H_N,x)],\qquad x\ge0.
\end{equation}
On the observation timescale the relevant value is $m_\sigma(\gamma t)$.
The cutoff keeps this quantity finite even when
$\mathcal M_\sigma=\infty$, and $\log(I+N)$ can replace $H_N$ at the
cost of fixed comparison factors.

\begin{proposition}[Truncated-moment bounds]\label{prop:truncatedmoment}
Let $c_0=1-\exp(-e^{-1})$ and
$d_\sigma(t)=\norm{\Phi_t\pi_0-\pi_0}_1$. For every $t>0$,
every phase rotation frequency, and every diagonal reset state,
\begin{equation}\label{eq:truncateddiagonal}
 \frac{2ae^{-a}c_0}{\gamma t}m_\sigma(\gamma t)
 \le d_\sigma(t)\le\frac{2a}{\gamma t}m_\sigma(\gamma t).
\end{equation}
For an arbitrary reset state, the same lower bound holds, and
\begin{equation}\label{eq:truncatedcoherent}
 d_\sigma(t)\le\frac{4a}{\gamma t}
 \left[m_\sigma(\gamma t)+\sqrt{m_\sigma(\gamma t)}\right].
\end{equation}
All constants are independent of the reset state.
\end{proposition}

The proof in Appendix~\ref{app:truncatedmoment} compares the truncated
extinction time of $n$ excitations with $\min(H_n,\gamma t)$ and then
uses positivity to bound coherences. For an infinite logarithmic moment,
$m_\sigma(x)\to\infty$, so these inequalities determine the rate up to
constants even with coherences. They hold uniformly across finite
cutoffs, whereas Theorem~\ref{thm:thermalmoment} supplies the exact
coefficient whenever the moment is finite.

The square-root term cannot simply be dropped uniformly over reset
states. For $|\psi_p\rangle=\sqrt{1-p}|0\rangle+\sqrt p|1\rangle$
and $\omega=0$, the moment is $\mathcal M_{\psi_p}=p$, whereas
$\gamma\norm{h_{\psi_p}}_1=2\sqrt{4p-3p^2}$
(Appendix~\ref{app:coherentreset}). A small coherence can therefore
dominate the population response near the vacuum.

The same truncated moment controls the stationary state, with the mean
time $1/r$ between resets in place of $t$. For pure loss,
\begin{equation}\label{eq:stationarytailorder}
 \norm{\rho_r^{\mathrm{ss}}-\pi_0}_1
 \asymp\frac r\gamma m_\sigma(\gamma/r)
 \qquad(r\downarrow0)
\end{equation}
for diagonal reset states, and also for arbitrary reset states with an
infinite logarithmic moment. The constants are independent of the state,
although for an infinite moment the onset may depend on it. For reset
states with coherences, the finite-rate upper bound retains a
square-root correction, as in \eqref{eq:truncatedcoherent};
Appendix~\ref{app:nesstruncated} gives the explicit bounds. Stationary
and finite-time errors are therefore governed by the tail at the same
scale, although their leading coefficients can differ
(Section~\ref{sec:weakreset}).

The tail comparison extends to every fixed finite temperature.
For $D=a(P_\sigma-\id)$ and vacuum input, an infinite logarithmic
moment gives
\begin{equation}\label{eq:thermaltailorder}
 \norm{\Phi_t\pi_0-\tau_{\bar n}}_1
 \asymp\frac{m_\sigma(\gamma t)}{\gamma t}
 \qquad(t\to\infty).
\end{equation}
The finite-time bounds in Appendix~\ref{app:thermaltruncated} include
an additive thermal term of order $1/t$ and an exponentially decaying
initial transient. Their constants are independent of the reset state
at fixed $a$ and $\bar n$; the onset in \eqref{eq:thermaltailorder}
may depend on its tail. Thermal noise therefore preserves the order
$m_\sigma(\gamma t)/(\gamma t)$, although this comparison does not
determine how it changes the coefficient.

\subsection{A logarithmic tail and its finite cutoffs}
An explicit case is
\begin{equation}\label{eq:logtailmain}
 \sigma=\sum_{k\ge1}\frac{6}{\pi^2k^2}\pi_{\lceil e^k\rceil},
 \qquad D=a(P_\sigma-\id).
\end{equation}
This is a normalized state with infinite logarithmic moment. Its weights
decrease quadratically in $k$, while its occupied excitation levels grow
exponentially. Levels with $k$ much larger than $\gamma s$ still retain
an excitation at time $s$. Their total weight makes the nonvacuum
population behave as $6/(\pi^2\gamma s)$.

The reset average in \eqref{eq:resetresponse} integrates this tail.
For vacuum input, the exact asymptotic is
\begin{equation}\label{eq:logratemain}
 \norm{\Phi_t\pi_0-\pi_0}_1
 \sim\frac{12a}{\pi^2\gamma}\frac{\log(1+\gamma t)}{t}.
\end{equation}
The calculation is given in Appendix~\ref{app:logarithmic}.
The perturbation is fixed and bounded, so Theorem~\ref{thm:strong}
still guarantees convergence. The extra logarithm comes from the time
integral of the nonvacuum population, which grows like $\log(\gamma t)$
instead of saturating. Other tails with an infinite moment give other
rates; the power-law family of Appendix~\ref{app:weakreset} decays as
$t^{-\beta}$. A continuous switch-on $a_t=a(1-e^{-\gamma t})$ gives the
same leading law and completes the family with $\Phi_0=\id$.

The state in \eqref{eq:logtailmain} also has infinite mean energy.
A finite cutoff changes the ultimate asymptote. Let
$\sigma_K=c_K\sum_{k=1}^K k^{-2}\pi_{\lceil e^k\rceil}$, where
$c_K=(\sum_{k=1}^K k^{-2})^{-1}$. Its logarithmic moment is
$\mathcal M_{\sigma_K}=c_K\log K+O(1)$ as $K\to\infty$.
For each fixed $K$ and every finite-energy input,
\begin{equation}\label{eq:cutoffrate}
 \norm{\Phi_t^{(K)}\rho-\pi_0}_1
 \sim\frac{2a\mathcal M_{\sigma_K}}{\gamma t},
\end{equation}
where $\Phi_t^{(K)}$ uses the reset target $\sigma_K$.
The highest occupied level has mean extinction time
$H_{\lceil e^K\rceil}/\gamma\sim K/\gamma$. The rescaled error can grow
logarithmically while $\gamma t\lesssim K$. Once the slowest occupied
levels have had time to relax, the rescaled error levels off and
\eqref{eq:cutoffrate} gives the $1/t$ asymptote.

The two asymptotic regimes depend on the order of the time and cutoff
limits. Figure~\ref{fig:tailcutoff} displays the increasing rescaled
error and the finite-cutoff plateaus. Appendix~\ref{app:logarithmic}
gives the asymptotic calculation; Proposition~\ref{prop:truncatedmoment}
provides bounds valid without taking either limit first.

\section{Cumulant dynamics and the Redfield distinction}\label{sec:cumulant}
\begin{figure*}[t]
\centering
\includegraphics[width=\textwidth]{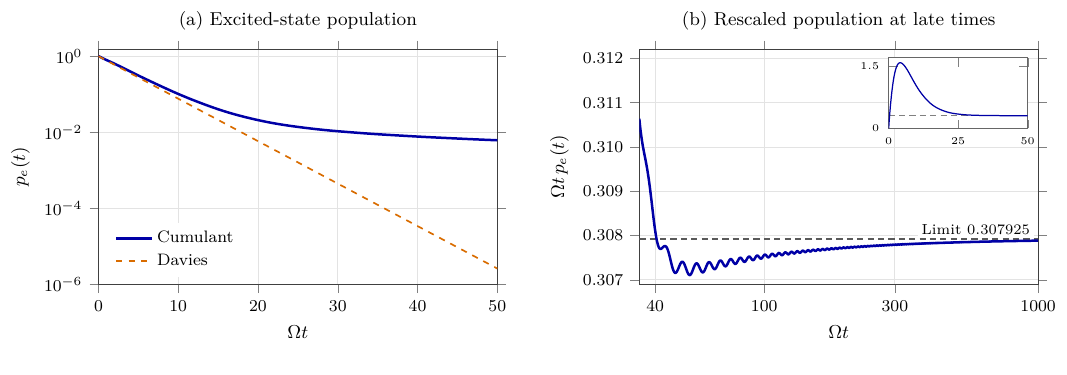}
\caption{An initially excited qubit in a vacuum Ohmic bath, with
$\omega_c=5\Omega$ and $\lambda^2\alpha=0.05$, corresponding to the
spectrum used in Fig.~1 of Ref.~\cite{Rivas2017}.
(a)~Excited population from the finite-time cumulant
[\eqref{eq:cumulantpopulation} and \eqref{eq:rivasspectral}] and from the
Davies semigroup, $p_e^{\rm D}(t)=\exp[-2\pi\lambda^2 J(\Omega)t]$.
(b)~Late-time cumulant population multiplied by $\Omega t$, with a
logarithmic time axis. The dashed line marks half the trace-norm
coefficient in \eqref{eq:rivasqubittail}. The inset shows the initial
transient of the same rescaled population on linear axes.}
\label{fig:rivasqubit}
\end{figure*}
The second-order cumulant map of a qubit coupled to a vacuum bath can
also produce an inverse-time tail without resetting. In the transverse model studied by
Rivas~\cite{Rivas2017}, the second-order cumulant's integrated
emission coefficient grows linearly in time, while its integrated
excitation coefficient approaches a positive constant. Their ratio
sets the late-time excited population.

This mechanism fits the exponent family in \eqref{eq:familyintro}.
In the interaction picture the second-order cumulant gives
\begin{equation}\label{eq:cumulantfamily}
 \widetilde\Phi_t=e^{\lambda^2K_2(t)},\qquad
 K_2(t)=tL_{\rm D}+R_t.
\end{equation}
Here $\lambda$ is the system--bath coupling and $L_{\rm D}$ is the
secular weak-coupling generator
\cite{Alicki1989,SchallerBrandes2008,Rivas2017,WinczewskiAlicki2021}.
When $L_{\rm D}$ relaxes to $\rho_*$ and $R_t$ is uniformly bounded,
\eqref{eq:leading} gives, at fixed $\lambda\ne0$,
\begin{equation}\label{eq:cumulantresponse}
\begin{split}
 \widetilde\Phi_t\rho-\rho_*
 ={}&-\frac1t(L_{\rm D}|_{\E})^{-1}R_t\rho_*\\
 &+O(t^{-2})+O(e^{-c\lambda^2t}),\qquad c>0.
\end{split}
\end{equation}
The estimate is uniform over inputs. A nonzero limiting forcing
$R_t\rho_*$ produces a nonzero coefficient.

For the vacuum Ohmic spectrum of Rivas,
$J(\nu)=\alpha\nu e^{-\nu/\omega_c}$ with $\omega_c=5\Omega$,
the two transition coefficients can be analyzed directly even though
the first absolute correlation moment diverges. Their finite-time
expressions~\cite{Rivas2017} and the population solution of
Ref.~\cite{SuarezEtAl2024} yield
\begin{equation}\label{eq:rivasqubittail}
 \lim_{t\to\infty}\Omega t\,\norm{\widetilde\Phi_t\rho-\pi_g}_1
 \simeq 0.6158491457 .
\end{equation}
Here $\Omega$ is the transition frequency and $\pi_g$ is the bare
ground state. The coherences decay exponentially, so the limit is
uniform over all qubit inputs at fixed nonzero coupling
(Appendix~\ref{app:rivasqubit}).

For an initially excited qubit, the state remains diagonal and the
error is exactly $2p_e(t)$. At $\Omega t=30$, the cumulant gives
$p_e=0.01065$ and the Davies semigroup gives $p_e=0.0004455$ for the
coupling in Fig.~\ref{fig:rivasqubit}. Figure~\ref{fig:rivasqubit}(b) shows how
$\Omega t\,p_e(t)$ approaches half the coefficient in
\eqref{eq:rivasqubittail}. Appendix~\ref{app:drudequbit} gives a
second example at positive temperature.

The direct population argument is useful when a first correlation
moment is unavailable. For a fixed finite-dimensional system with
finitely many couplings to a centered stationary reservoir,
integrable correlations $C_{\alpha\beta}(s)$ give $R_t=o(t)$.
A finite first absolute moment for each pair guarantees bounded
$R_t$ and hence \eqref{eq:cumulantresponse}. Appendix~\ref{app:cumulant}
derives these conditions for the full cumulant, including the Lamb
shift. The bounded oscillatory terms are also visible in
Ref.~\cite[Appendix D, Eq.~(D172)]{WinczewskiAlicki2021}. For a
thermal reservoir and a relaxing Davies generator, $\rho_*$ is the
Gibbs state of the reference Hamiltonian. It is invariant under
the free rotation, so the error norm is the same in both pictures.

The population tail concerns the ordinary cumulant. The corresponding
finite-upper-limit Bloch--Redfield equation
\cite{Bloch1957,Redfield1957} uses $\lambda^2\dot K_2(t)$ as its
instantaneous generator and yields a time-ordered propagator.
The two constructions agree to second order at fixed time, but this
does not identify their long-time limits, so the inverse-time tail
belongs to the ordinary exponential. An autonomous nonsecular Redfield
generator changes the Davies exponent by a term linear in $t$. When it
has a relaxing, stable trace-zero part, it approaches its own stationary
vector exponentially (Appendix~\ref{app:cumulant}). The fully secular,
relaxing Davies equation is the case $D_t=0$ and also relaxes
exponentially.

The exact microscopic dynamics also has a different equilibrium
reference. At sufficiently small coupling the vacuum spin--boson
model relaxes, on system observables, to the interacting ground state
\cite[Theorem~1.1 and Remark~1.2]{DeRoeckKupiainen2013}, which exists
and depends analytically on the coupling
\cite[Theorem~1]{HaslerHerbst2011}. Its bare excited population is
positive at second order. For the spectrum and coupling in
Fig.~\ref{fig:rivasqubit}, the second-order term evaluates to about
$0.040$ (Appendix~\ref{app:rivasqubit}). This is already larger than
the cumulant prediction at $\Omega t=30$. It provides a scale for
the equilibrium shift, while \eqref{eq:rivasqubittail} quantifies the
approach to $\pi_g$ within the cumulant map.

At positive temperature, resonance theory gives complementary
microscopic estimates. After separating stationary and resonance
contributions, K\"onenberg and Merkli bound the remaining propagator
matrix elements by $O(t^{-1})$~\cite[Theorem~1.1]{KonenbergMerkli2016}.
For thermal Bose reservoirs, Merkli bounds the error of the usual
Davies semigroup uniformly in time by $O(\lambda^2)$ under analyticity
\cite[Sec.~2.8]{Merkli2020Resonance}, and by $O(|\lambda|^{1/4})$
under weaker regularity~\cite{Merkli2022Oscillation,Merkli2022Markovian}.
These time-uniform estimates control the error at small coupling.
With stronger regularity a renormalized semigroup can preserve the
interacting equilibrium and approximate the microscopic dynamics with
an error that decays at late times
\cite{KonenbergMerkli2017,KonenbergMerkli2019,Merkli2020Resonance}.

\section{Discussion}\label{sec:discussion}
In the oscillator, the time to remove an excitation depends on how
high in the ladder it starts. The logarithmic moment measures this
accumulated relaxation and sets the threshold for a finite response
to weak resetting. The truncated moment then accounts for the
portion of that response sampled during a finite protocol. This is
why a finite cutoff recovers a $1/t$ tail only after the highest
populated levels have had time to relax.

The timescale in this statement is the observation duration along a
family with reset rate $a_t/t$. At a fixed positive reset rate the
approach to the perturbed stationary state is exponential
(Appendix~\ref{app:weakreset}). Power laws can also occur in resetting
with memory~\cite{BoyerMajumdar2024,DeMauroKulkarniMajumdar2026}.
The duration-dependent reset rate considered here gives a direct
accuracy constraint for a preparation protocol.

\subsection{Ancilla preparation and physical resources}\label{sec:ancilla}
The duration a dissipative preparation needs to reach a given accuracy
follows directly from Sec.~\ref{sec:example}. For the one-photon input,
the trace distance to the target vacuum is
\[
 d(t)=\tfrac12\norm{\Phi_t\pi_1-\pi_0}_1
 =e^{-\gamma t-a_t}+p_t\sim\frac{a}{\gamma t}.
\]
Let $t_\epsilon$ be the duration beyond which the error stays below
the tolerance $\epsilon$. At fixed $a>0$, as $\epsilon\downarrow0$,
\[
 t_\epsilon\sim\frac{a}{\gamma\epsilon},
 \qquad t_\epsilon^{(0)}=\frac1\gamma\log\frac1\epsilon,
\]
where $t_\epsilon^{(0)}$ is the duration for unperturbed pure loss.
Here the error comes from residual population. In the
vacuum--one-photon sector the off-diagonal element is attenuated in
modulus by $e^{-\gamma t/2-a_t}$. For $H=\hbar\omega N$ with $\omega>0$,
the residual energy for the one-photon input is $\hbar\omega d(t)$.

For an independently prepared ancilla, contractivity carries its
trace-distance error into an upper bound on the change in the output
of the subsequent circuit. The preparation duration can therefore
enter syndrome-extraction schedules and estimates of the number of
parallel preparation units
\cite{GeherEtAl2025,FellousAsianiEtAl2023}. A full energy accounting also
requires a physical model of preparation and recovery of auxiliary
devices, including reset, control, and cooling
\cite{HorodeckiEtAl2025Energy,YehiaEtAl2024,Auffeves2022,Ezratty2025}.
The relation above supplies the time needed to meet the specified
preparation tolerance.

\subsection{Outlook}
For the oscillator, natural extensions are correlated multimode reset
targets, coupled bosonic generators, higher-order response, and bounds
uniform in temperature; unbounded perturbations would require
additional generation and domain estimates. For the microscopic qubit,
the open question is the leading decay law after subtracting the
interacting equilibrium. To identify a $1/t$ coefficient from an effective description,
one must subtract the respective equilibria and show that the two
transients differ by $o(t^{-1})$.

\begin{acknowledgments}
\interlinepenalty=10000\relax
M.~W. acknowledges support from the National Science Centre, Poland,
under OPUS~25 Grant No.~2023/49/B/ST2/02468.

The author used ChatGPT (OpenAI, GPT-6 Astra, October 2026) for assistance with literature searches, language and structural editing, and, to a limited extent, mathematical derivations. Claude (Anthropic, Claude Opus 5.5, November 2026) was used to check the correctness of the mathematical arguments, numerical results, and cited findings, and to assist with language and structural editing. The author takes full responsibility for the scientific content and the final text.
\end{acknowledgments}

\section*{Code availability}
Code for the figures and numerical checks is available from the author
upon request.

\appendix
\section{Trace-class perturbation estimates}\label{app:traceclass}
\subsection{Framework and generator domains}\label{app:framework}
Let $\E_h$ be the real subspace of Hermitian elements of
$\E=\ker\tr$. The restricted semigroup has generator
\begin{equation}\label{eq:domainA}
 A=L|_\E,\qquad \dom A=\dom L\cap\E.
\end{equation}
For each fixed $t>0$, the bounded perturbation theorem
\cite[Theorem III.1.3]{EngelNagel2000} gives
\begin{equation}\label{eq:domainG}
 G_t=L+D_t/t,\qquad\dom G_t=\dom L.
\end{equation}
Thus $e^{tL+D_t}$ denotes the semigroup value $e^{tG_t}$; no power
series for an unbounded operator is required. In finite dimension,
the unperturbed generator has the GKSL form
\cite{Gorini1976,Lindblad1976}. Standard background is given in
Refs.~\cite{Wolf2012,EngelNagel2000}.

The channels $T_s$ are contractions in the induced trace norm. Condition
\eqref{eq:relax} implies $T_sX\to PX$ for every $X\in\X$, since every
trace-class operator is a linear combination of positive trace-class
operators. The semigroup property gives $T_s\rho_*=\rho_*$, so that
$\rho_*\in\dom L$ and $L\rho_*=0$.

The Dyson--Phillips series preserves Hermiticity term by term.
Taking the trace of any term of positive order gives zero, since
the leftmost perturbation is trace-annihilating and the intervening
unperturbed maps preserve trace. Thus the perturbed semigroup is
trace preserving. These arguments do not establish positivity.

\subsection{Strong relaxation}\label{app:strong}
\begin{proof}[Proof of Theorem~\ref{thm:strong}]

Let $U_t(s)=e^{s(A+D_t|_\E/t)}$. Contractivity of $T_s|_\E$ and the
bounded-perturbation estimate give $\norm{U_t(s)}\le e^{Ks/t}$.
The triangular decomposition $\X=\mathbb C\rho_*\oplus\E$ gives the
mild variation-of-constants formula
\begin{equation}\label{eq:mild}
 \Phi_t\rho-\rho_*=U_t(t)x+\frac1t\int_0^tU_t(s)b_t\,ds.
\end{equation}
All integrals are strong Bochner integrals. In particular, no domain
assumption on $x$ or $b_t$ is needed. Duhamel's formula in the order
\[
 U_t(s)y=T_sy+\frac1t\int_0^sU_t(s-u)D_tT_uy\,du,
 \qquad y\in\E,
\]
implies
\[
 \norm{U_t(t)x}_1\le\norm{T_tx}_1+
 \frac{Ke^K}{t}\int_0^t\norm{T_ux}_1\,du.
\]
Applying the same identity to $b_t$ and interchanging the two integrations
in the resulting norm bound yields
\begin{equation}\label{eq:forcedestimate}
 \norm{\frac1t\int_0^tU_t(s)b_t\,ds}_1
 \le\frac1t\int_0^t e^{K(1-u/t)}\norm{T_ub_t}_1\,du.
\end{equation}
Indeed, the coefficient of $\norm{T_ub_t}_1/t$ is
$1+(K/t)\int_u^t e^{K(s-u)/t}ds=e^{K(1-u/t)}$; the case $K=0$
follows directly. This proves \eqref{eq:cesaroestimate}.

For fixed $x\in\E$, strong relaxation and the Ces\`aro mean theorem imply
that the first two terms tend to zero. A uniformly bounded family of
operators converging strongly to zero converges uniformly on each
norm-compact set, by a finite-net argument. Applying this fact to $T_s|_\E$
on the compact closure of $\{b_t\}$ proves \eqref{eq:forcingcesaro}.
\end{proof}

\subsection{Compactness from an energy bound}\label{app:compact}
Let $V\ge I$ be self-adjoint with compact resolvent. A sufficient
condition for relative compactness of the forcing set is
\begin{equation}\label{eq:weightedforcing}
 D_t\rho_*=V^{-1/2}Y_tV^{-1/2},\qquad
 \sup_t\norm{Y_t}_1<\infty.
\end{equation}

Under \eqref{eq:weightedforcing}, the operators $Y_t$ form a bounded
set in trace class. The operator $V^{-1/2}$ is compact.
To prove compactness,
approximate the compact operator $V^{-1/2}$ in operator norm by finite-rank
operators. The corresponding two-sided sandwiches approximate the map
$Y\mapsto V^{-1/2}YV^{-1/2}$ in the induced trace norm and have
finite-dimensional range. This also explains why energy control may replace
uniform relaxation on the entire state space.

\subsection{Uniform exponential relaxation}\label{app:uniform}
\begin{theorem}[Displacement and transient]\label{thm:main}
Assume \eqref{eq:structure}, \eqref{eq:basebound}, and
\begin{equation}\label{eq:sublinear}
 \norm{D_t}/t\longrightarrow0.
\end{equation}
Then $\Phi_t\to P$ in induced trace norm. For sufficiently large $t$,
$B_t=A+D_t|_\E/t$ is invertible. Define
\begin{equation}\label{eq:drift}
 v_t=-\frac1tB_t^{-1}D_t\rho_*,\qquad
 \rho_t^{\mathrm{fr}}=\rho_*+v_t.
\end{equation}
The vector $\rho_t^{\mathrm{fr}}$ is the unique trace-one kernel vector
of $L+D_t/t$, and for every state $\rho$,
\begin{equation}\label{eq:exact}
 \Phi_t\rho-\rho_*=v_t+e^{tB_t}(\rho-\rho_*-v_t).
\end{equation}
If, more strongly, $\sup_t\norm{D_t}<\infty$, then uniformly over all
initial states,
the expansion \eqref{eq:leading} holds.
\end{theorem}

Fix $\gamma,M$ as in \eqref{eq:basebound}, and define
\begin{equation}\label{eq:definitions}
\begin{gathered}
 R_t=D_t/t,\qquad B_t=A+R_t|_\E,\\
 r_t=\norm{R_t|_\E},\qquad\beta_t=\gamma-Mr_t,\\
 f_t=\norm{D_t\rho_*}_1.
\end{gathered}
\end{equation}
For all sufficiently large $t$, $\beta_t\ge\gamma/2>0$.

\begin{proposition}[Quantitative uniform bounds]\label{prop:uniformtechnical}
Under Theorem~\ref{thm:main}, for sufficiently large $t$,
\begin{equation}\label{eq:stability}
 \norm{e^{sB_t}}\le M e^{-\beta_t s},\qquad
 \norm{B_t^{-1}}\le M/\beta_t\quad(s\ge0).
\end{equation}
The vector $v_t$ in \eqref{eq:drift} belongs to $\dom A$,
and $\rho_t^{\mathrm{fr}}\in\dom L$ is the unique trace-one kernel vector
of $L+D_t/t$. For every density operator $\rho$,
\begin{equation}\label{eq:upper}
\begin{split}
 \norm{\Phi_t\rho-\rho_*}_1
 &\le 2M e^{-\beta_t t}\\
 &\quad+\frac{M(1-e^{-\beta_t t})}{\beta_t}\frac{f_t}{t}.
\end{split}
\end{equation}
The same right-hand side bounds $\norm{\Phi_t-P}_{1\to1}$.
\end{proposition}

\begin{proof}[Proof of Theorem~\ref{thm:main} and Proposition~\ref{prop:uniformtechnical}]

For a fixed $t$, Duhamel's formula gives
\[
 e^{sB_t}=e^{sA}+\int_0^s e^{(s-u)A}R_t|_\E e^{uB_t}\,du.
\]
Multiplying the norm inequality by $e^{\gamma s}$ and applying Gronwall
gives $\norm{e^{sB_t}}\le M e^{-(\gamma-Mr_t)s}$. Integration of this stable
semigroup proves $B_t^{-1}x=-\int_0^\infty e^{sB_t}x\,ds$ and the inverse
bound. The integral belongs to $\dom B_t=\dom A$: this follows first for
finite integration intervals from the generator identity, and then from
closedness of $B_t$. Exponential stability also implies
$\ker B_t=\{0\}$, completing the inverse construction. Thus the formula
is also valid when $A$ is unbounded.

For $X=a\rho_*+x$, where $a=\tr X$ and $x\in\E$, the frozen operator
$G_t=L+R_t$ has the block form
\begin{equation}\label{eq:block}
 G_t=\begin{pmatrix}0&0\\ R_t\rho_*&B_t\end{pmatrix}.
\end{equation}
Its kernel is spanned by $\rho_t^{\mathrm{fr}}$ in \eqref{eq:drift}. On the invariant
subspace $\E$, its exponential is $e^{tB_t}$, which proves \eqref{eq:exact}.
Equivalently, the exact variation-of-constants formula is
\begin{equation}\label{eq:variation}
 \Phi_t X-PX=e^{tB_t}(X-\rho_*\tr X)
 +(\tr X)\int_0^t e^{sB_t}\frac{D_t\rho_*}{t}\,ds.
\end{equation}
Since $\norm{X-\rho_*\tr X}_1\le2\norm X_1$ and
$|\tr X|\le\norm X_1$, integration of \eqref{eq:stability} proves the upper
bound both on states and on all trace-class operators. Finally
$f_t/t\le\norm{D_t}/t\to0$. The leading expansion in
Theorem~\ref{thm:main} follows from the response estimates
proved in Appendix~\ref{app:response}.
\end{proof}
\begin{remark}[Limiting state and frozen invariant vector]\label{rem:fixed}
The limit is $\rho_*$, whereas the unique trace-one fixed vector of
$\Phi_t$ for sufficiently large $t$ is $\rho_t^{\mathrm{fr}}$.
This vector is positive if $\Phi_t$ is CPTP. Under
Theorem~\ref{thm:main},
\[
 \Phi_t\rho_*=\rho_*\quad\Longleftrightarrow\quad D_t\rho_*=0.
\]
Thus a common asymptotic state need not be a common fixed point.
\end{remark}
\paragraph{Justification of Remark~\ref{rem:fixed}.}
Iteration of a fixed CPTP map $\Phi_t$ converges, by
\eqref{eq:exact}, to $X\mapsto\tr(X)\rho_t^{\mathrm{fr}}$.
The limit must be positive, so $\rho_t^{\mathrm{fr}}$ is a density operator.
Exponential stability makes $I-e^{tB_t}$ invertible. Together with
\eqref{eq:exact} and \eqref{eq:drift}, this proves the fixed-point
equivalence in Remark~\ref{rem:fixed}.

\subsection{Integrated response and worst-case bounds}\label{app:response}
\begin{corollary}[Integrated response]\label{cor:response}
Under the assumptions of Theorem~\ref{thm:main}, put
$k=\norm{A^{-1}}$, $h_t=\norm{A^{-1}D_t\rho_*}_1$, and let $r_t$ be as in
\eqref{eq:definitions}. Whenever $kr_t<1$,
\begin{align}
 \frac{h_t}{t(1+kr_t)}&\le\norm{v_t}_1
 \le\frac{h_t}{t(1-kr_t)},\label{eq:response}\\
 \norm{v_t+\frac1tA^{-1}D_t\rho_*}_1
 &\le\frac{kr_t}{1-kr_t}\frac{h_t}{t}.
 \label{eq:relativeremainder}
\end{align}
If $\sup_t\norm{D_t}<\infty$ and $D_t\rho_*\to b\ne0$, then
$t\norm{\Phi_t\rho-\rho_*}_1\to\norm{A^{-1}b}_1>0$ for every state $\rho$.
\end{corollary}

The response also gives a two-sided description of the worst-case error.
Define $W(t)=\sup_\rho\norm{\Phi_t\rho-\rho_*}_1$ and
\begin{equation}\label{eq:epsilon}
 \varepsilon_t=\norm{e^{tB_t}|_{\E_h}}_{1\to1}.
\end{equation}
The norm in \eqref{eq:epsilon} is on the real Hermitian trace-zero subspace.
Once $\varepsilon_t\le1/2$,
\begin{equation}\label{eq:responseworst}
 \frac{\norm{v_t}_1+\varepsilon_t}{3}
 \le W(t)\le2(\norm{v_t}_1+\varepsilon_t).
\end{equation}
Consequently, $W(t)\asymp h_t/t+\varepsilon_t$. The two terms describe,
respectively, the displacement of the frozen invariant vector and the
remaining transient. 
For unbounded $A$, $h_t$ need not be bounded below by a constant times
$\norm{D_t\rho_*}_1$.

\paragraph{Proof of the response estimates.}

The inverse formula $A^{-1}x=-\int_0^\infty T_sx\,ds$ follows from
\eqref{eq:basebound}, including when $A$ is unbounded.
Applying $A^{-1}$ to $B_tv_t=-D_t\rho_*/t$ gives
\[
 (I+A^{-1}D_t|_\E/t)v_t=-A^{-1}D_t\rho_*/t.
\]
The Neumann series bounds the inverse by $(1-kr_t)^{-1}$;
the triangle inequality applied before inversion gives the lower bound
in \eqref{eq:response}. Subtracting the zeroth-order term gives
\eqref{eq:relativeremainder}. If $K=\sup_t\norm{D_t}<\infty$, the
remainder is $O(t^{-2})$. Since $tr_t\le K$, \eqref{eq:stability} gives
$\norm{e^{tB_t}}\le Me^{MK}e^{-\gamma t}$. Combining these estimates
with \eqref{eq:exact} proves
\eqref{eq:leading}, uniformly over states, and hence the stated limit
when $D_t\rho_*\to b$.

For the worst-case estimate, use the notation in \eqref{eq:epsilon}.
The vector $v_t$ is Hermitian, since $D_t\rho_*$ is Hermitian and the
inverse integral preserves Hermiticity.
For $X\in\E_h$ with $\norm X_1=1$, its Jordan decomposition gives
$X=(\rho_+-\rho_-)/2$ for states $\rho_\pm$. Thus $W(t)\ge\varepsilon_t$.
Using the initial state $\rho_*$ in \eqref{eq:exact} also gives
$W(t)\ge(1-\varepsilon_t)\norm{v_t}_1$, whereas the same identity implies
$W(t)\le(1+\varepsilon_t)\norm{v_t}_1+2\varepsilon_t$. Once
$\varepsilon_t\le1/2$, these inequalities give \eqref{eq:responseworst}.
Combining it with \eqref{eq:response} yields
$W(t)\asymp h_t/t+\varepsilon_t$.

\section{Finite-dimensional spectral estimates}\label{app:finite}\label{sec:finite}
Throughout this section and the following finite-dimensional examples,
$\mathcal H=\mathbb C^d$, $2\le d<\infty$. A strongly continuous quantum
Markov semigroup then has a bounded GKSL generator. The symbol $\Tr$
denotes the trace of a linear map, as opposed to the matrix trace $\tr$.

\subsection{Spectral criterion}\label{app:spectral}
\begin{proposition}\label{prop:spectral}
Condition \eqref{eq:relax} holds if and only if $0$ is an algebraically
simple eigenvalue of $L$, with stationary density matrix $\rho_*$, and all
other eigenvalues have strictly negative real part. Equivalently, $A$ is
Hurwitz. If $g=-\max_{z\in\spec(A)}\Re z>0$, then for every
$0<\gamma<g$ there is $M\ge1$ such that \eqref{eq:basebound} holds.
\end{proposition}
Thus no diagonalizability or detailed balance is required. The exact gap
$g$ need not admit a bound $Me^{-gs}$ if a slow Jordan block is present.

\begin{proof}[Proof of Proposition~\ref{prop:spectral}]

Density matrices span $M_d$, so convergence on states implies convergence
$e^{sL}\to P$ in operator norm in finite dimension. The decomposition
$M_d=\mathbb C\rho_*\oplus\E$ then reduces the assertion to the Jordan
normal form of $A$. All its eigenvalues must have negative real part;
conversely, this condition gives convergence and algebraic simplicity of
the zero eigenvalue of $L$. Polynomial factors from Jordan blocks are
absorbed by $e^{(g-\gamma)s}$, giving \eqref{eq:basebound}.
\end{proof}
\subsection{Individual-state estimates}\label{app:finiterates}
Finite dimension allows us to express the response in terms of
$f_t=\norm{D_t\rho_*}_1$. Since $D_t\rho_*=-tB_tv_t$,
\begin{equation}\label{eq:driftbounds}
 \frac{f_t}{t(\norm A+r_t)}
 \le\norm{v_t}_1\le\frac{M f_t}{\beta_t t}.
\end{equation}
Put $\delta_t=M e^{-\beta_t t}$. The exact decomposition
\eqref{eq:exact} and $\norm{\rho-\rho_*}_1\le2$ give
\begin{equation}\label{eq:individualbounds}
\begin{aligned}
 \norm{\Phi_t\rho-\rho_*}_1
 &\ge\max\{0,(1-\delta_t)\norm{v_t}_1-2\delta_t\},\\
 \norm{\Phi_t\rho-\rho_*}_1
 &\le(1+\delta_t)\norm{v_t}_1+2\delta_t.
\end{aligned}
\end{equation}
For the input $\rho_*$ these sharpen to
\begin{equation}\label{eq:equilibriumbounds}
 (1-\delta_t)\norm{v_t}_1
 \le\norm{\Phi_t\rho_*-\rho_*}_1
 \le(1+\delta_t)\norm{v_t}_1.
\end{equation}
If $D_t\rho_*=0$, the error decays exponentially. If $D_t$ is bounded
and $f_t$ is eventually bounded away from zero, it is $\Theta(t^{-1})$
uniformly over all states. A lower bound covering vanishing forcing
must instead concern the worst-case error
\begin{equation}\label{eq:worst}
 W(t)=\sup_{\rho\text{ a state}}\norm{\Phi_t\rho-\rho_*}_1,
\end{equation}
since $D_t=0$ and $\rho=\rho_*$ give zero error for an individual input.

\subsection{Two-sided spectral bounds}\label{app:twosided}
\begin{theorem}[Two-sided worst-case bounds]\label{thm:twosided}
Let $L$ be relaxing and assume \eqref{eq:structure} and \eqref{eq:sublinear}.
For every $0<\eta<g$ there are
$c_\eta,C_\eta>0$ and $T_\eta$ such that for $t\ge T_\eta$,
\begin{equation}\label{eq:twosided}
\begin{aligned}
 W(t)&\ge c_\eta\left(\frac{f_t}{t}+e^{-(g+\eta)t}\right),\\
 W(t)&\le C_\eta\left(\frac{f_t}{t}+e^{-(g-\eta)t}\right).
\end{aligned}
\end{equation}
If $\norm{D_t}\le K$, an additional explicit lower bound, valid for each
$t$ for which this bound on $D_t$ holds, is
\begin{equation}\label{eq:detbound}
\begin{aligned}
 W(t)&\ge\exp\left(\frac{t\Tr A+\Tr(D_t|_{\E_h})}{d^2-1}\right)\\
 &\ge\exp\left(\frac{t\Tr A}{d^2-1}-K\right).
\end{aligned}
\end{equation}
Here $\Tr A$ can equivalently be computed on $\E_h$ or on its complexification.
\end{theorem}

These bounds give an exponential lower envelope and a $1/t$ upper envelope
for the bounded-perturbation class. There is no universal polynomial lower
bound. The leading term in \eqref{eq:leading} gives much more information
when the action of $D_t$ on $\rho_*$ is known.

\begin{proof}[Proof of Theorem~\ref{thm:twosided}]

For $X\in\E_h$ with $\norm X_1=1$, the positive and negative parts give
$X=(\rho_+-\rho_-)/2$ for states $\rho_\pm$. Hence
\[
 \norm{e^{tB_t}X}_1
 =\tfrac12\norm{\Phi_t\rho_+-\Phi_t\rho_-}_1\le W(t).
\]
Every induced real norm bounds the spectral radius of its complexification.
Consequently $W(t)\ge e^{t s(B_t)}$, where $s(B)=\max\Re\spec B$.
Eigenvalue continuity in finite dimension and $B_t\to A$ imply
$s(B_t)\to-g$, so eventually $W(t)\ge e^{-(g+\eta)t}$.
Equations \eqref{eq:driftbounds} and \eqref{eq:equilibriumbounds} also imply
$W(t)\ge c f_t/t$ for some $c>0$ eventually. The maximum of these two lower
bounds bounds a positive constant times their sum. For the upper bound,
choose $g-\eta<\gamma<g$ and then $t$ so large that
$\beta_t\ge g-\eta$ in \eqref{eq:upper}.

Finally, an operator on the $m=d^2-1$ dimensional real space $\E_h$ has norm
at least $|\det T|^{1/m}$. Apply this to
$T=e^{tA+D_t|_{\E_h}}$ and use
$|\Tr(D_t|_{\E_h})|\le m\norm{D_t|_{\E_h}}\le mK$.
\end{proof}

\section{Oscillator estimates and exact solutions}\label{app:oscillator}
\subsection{Relaxation under an energy constraint}\label{app:oscenergy}
\begin{lemma}[Energy-constrained relaxation]\label{prop:oscenergy}
Put $C_E=\pi\sqrt{(2\bar n+1)E+\bar n}$.
For every $s\ge0$ and every $\rho\in\mathcal S_E$,
\begin{align}\label{eq:energybound}
 \norm{T_s\rho-\tau_{\bar n}}_1
 &\le\min\left\{2,\frac{2C_E}{\pi}
 \arcsin(e^{-\gamma s/2})\right\}\notag\\
 &\le C_E e^{-\gamma s/2}.
\end{align}
The semigroup relaxes strongly on the entire trace class, but
the uniform separation \eqref{eq:oscnotuniform} holds.
\end{lemma}

\begin{proof}[Proof of Lemma~\ref{prop:oscenergy}]

Couple the input mode to an environment mode $\hat b$ in the thermal state
$\tau_{\bar n}$ through a beam splitter. Its Hermitian generator is
$H=i(\hat a^\dagger\hat b-\hat a\hat b^\dagger)$, and the difference between the mixing
angle at transmissivity $e^{-\gamma s}$ and that at zero transmissivity is
$\delta_s=\arcsin(e^{-\gamma s/2})$. On the product of the input state and
the thermal environment,
\[
 \langle H^2\rangle=(2\bar n+1)\tr(N\rho)+\bar n.
\]
The terms containing $\hat b^2$ or $(\hat b^\dagger)^2$ vanish.
Here $\langle H^2\rangle$ denotes the quadratic form
$\norm{H\psi}^2$ on a product purification $\psi$.
Its finiteness, obtained by number-state truncation and closedness of $H$,
puts $\psi$ in $\dom H$. Consequently,
$\norm{(e^{-i\delta_sH}-I)\psi}\le\delta_s\norm{H\psi}$.
The trace distance between pure states is at most twice their vector
distance. Contractivity under the partial trace gives the first bound
in \eqref{eq:energybound}; at zero transmissivity the system output is
$\tau_{\bar n}$. The rotation preserves both $N$ and $\tau_{\bar n}$.
Finally, $\arcsin x\le\pi x/2$ on $[0,1]$ gives the second bound.

Every state can be approximated in trace norm by normalized finite-energy
truncations. Contractivity and \eqref{eq:energybound} therefore prove
strong relaxation without any energy assumption on a fixed initial state.
For the opposite, uniform statement, use coherent inputs with amplitudes
of arbitrarily large modulus. At fixed $s<\infty$, a measured output
quadrature has a mean of arbitrarily large modulus and a bounded variance,
whereas the equilibrium quadrature distribution has zero mean and finite
variance. Threshold measurements distinguish these states with error
tending to zero. This proves \eqref{eq:oscnotuniform}. At zero temperature
one can also use the exact expression
$2\sqrt{1-e^{-e^{-\gamma s}|\alpha|^2}}$ for a coherent input $|\alpha\rangle$.
\end{proof}
\subsection{Integrated relaxation with coherences}\label{app:logintegrability}
\begin{lemma}[Integrated relaxation]\label{lem:logintegrability}
For pure loss, $\bar n=0$, and an arbitrary density operator $\sigma$,
\begin{equation}\label{eq:integratedmoment}
 \frac{2\mathcal M_\sigma}{\gamma}
 \le \int_0^\infty\norm{T_s\sigma-\pi_0}_1\,ds
 \le \frac4\gamma\bigl(\mathcal M_\sigma+\sqrt{\mathcal M_\sigma}\bigr).
\end{equation}
The upper bound applies when $\mathcal M_\sigma<\infty$; the integral
diverges when $\mathcal M_\sigma=\infty$.
\end{lemma}

\begin{proof}[Proof of Lemma~\ref{lem:logintegrability}]
Set $Q=I-\pi_0$, $p_n=\langle n|\sigma|n\rangle$, and
$q(s)=\tr(QT_s\sigma)$. Pure loss gives
\[
 q(s)=\sum_{n\ge0}p_n[1-(1-e^{-\gamma s})^n],
 \qquad
 \int_0^\infty q(s)ds=\frac{\mathcal M_\sigma}{\gamma},
\]
where the integral identity follows from Tonelli and
\eqref{eq:harmonicresponse}. A vacuum measurement yields
$\norm{T_s\sigma-\pi_0}_1\ge2q(s)$.

For the upper bound decompose the positive operator $\sigma_u=T_u\sigma$ as
\[
 \sigma_u=(1-q(u))\pi_0+|v_u\rangle\langle0|
       +|0\rangle\langle v_u|+B_u,
\]
with $v_u\perp|0\rangle$, $B_u=Q\sigma_uQ\ge0$, and $\tr B_u=q(u)$.
Positivity implies $\norm{v_u}^2\le q(u)(1-q(u))$.
Vacuum coherences evolve exactly as
\[
 T_r(|v\rangle\langle0|)
 =|e^{-(\gamma/2+i\omega)Nr}v\rangle\langle0|.
\]
This identity follows on number-state matrix units and extends to all
$v\perp|0\rangle$ by continuity. Hence their combined trace norm is
at most $2e^{-\gamma r/2}\norm v$. Also
$\norm{T_rB_u-q(u)\pi_0}_1\le2q(u)$.
Choosing $u=r=s/2$ gives
\begin{equation}\label{eq:coherencehalftime}
 \norm{T_s\sigma-\pi_0}_1
 \le2q(s/2)+2e^{-\gamma s/4}\sqrt{q(s/2)}.
\end{equation}
The first term integrates to $4\mathcal M_\sigma/\gamma$.
For the second, Cauchy--Schwarz gives
\[
 2\int_0^\infty e^{-\gamma s/4}\sqrt{q(s/2)}ds
 \le\frac{4\sqrt{\mathcal M_\sigma}}{\gamma}.
\]
Together with the vacuum-measurement lower bound, this proves the claim.
\end{proof}

\subsection{General logarithmic-moment forcing}\label{app:oscforcing}
\begin{corollary}[Logarithmic-moment forcing]\label{cor:oscforcing}
Let $0\le\bar n<\infty$. Let $D_t$ be Hermiticity-preserving and
trace-annihilating, with $\sup_{t\ge t_0}\norm{D_t}\le K$. Suppose the logarithmic-moment condition \eqref{eq:oscforcingmoment} holds.
With $C_E$ as in Lemma~\ref{prop:oscenergy}, for $t\ge t_0$,
\begin{equation}\label{eq:oscforcingbound}
\begin{split}
 \sup_{\rho\in\mathcal S_E}\norm{\Phi_t\rho-\tau_{\bar n}}_1
 &\le C_E e^{-\gamma t/2}\\
 &\quad+\frac{2e^K}{\gamma t}
       \bigl(KC_E+2F_H+2\sqrt{KF_H}\bigr).
\end{split}
\end{equation}
\end{corollary}
The condition is sufficient for general signed forcing; necessity is
asserted only for fixed reset targets. Positivity of $\Phi_t$ is a
separate requirement.
For comparison with Ref.~\cite{Salzmann2024}, take $\bar n=0$, a fixed $D$,
$\omega=0$, and assume the stronger condition $\tr(N|D\pi_0|)<\infty$.
This is equivalent to finite first particle-number moments of the positive
and negative parts of $D\pi_0$, as in its Remark~5.7.
The attenuator in Ref.~\cite[Eq.~(5.25)]{Salzmann2024} is
parametrized by transmissivity $e^{-t}$. In this convention our
family has damping strength $\Gamma=\gamma t$ at unit observation
duration.
Thus \eqref{eq:oscforcingbound} gives $O(\Gamma^{-1})$ where its
Eq.~(5.34) gives $O((\log\Gamma)/\Gamma)$. The order is attained by
the fixed one-photon reset $D=a(P_{\pi_1}-\id)$, $a>0$: setting $a_t=a$
in \eqref{eq:oscreset} gives the asymptote $2a/(\gamma t)$ for a vacuum
input. The continuous switch-on version is given in
Appendix~\ref{app:exactinverse}. The constant in \eqref{eq:oscforcingbound}
is not claimed to be optimal. Proposition~5.6 assumes its perturbation
$\mathcal L\in\mathcal B(\mathfrak S_1)$; only the damping generator
may be unbounded. Its explicit energy-dependent right-hand side also
gives uniformity on each $\mathcal S_E$ for a fixed admissible
perturbation. Our comparison is between these common classes, and
does not address the discrete Zeno product.

\begin{proof}[Proof of Corollary~\ref{cor:oscforcing}]
Write $b_t=D_t\tau_{\bar n}=b_{t,+}-b_{t,-}$ and
$m_t=\tr b_{t,+}=\tr b_{t,-}=\norm{b_t}_1/2$.
First consider the pure-loss semigroup $T_s^{(0)}$ with the same
$\gamma,\omega$. For $m_t>0$, apply Lemma~\ref{lem:logintegrability} to
$b_{t,+}/m_t$ and $b_{t,-}/m_t$. The vacuum terms cancel. With
$F_{t,\pm}=\tr(H_Nb_{t,\pm})$, the integrated bound is
\begin{align*}
 \int_0^\infty\norm{T_s^{(0)}b_t}_1ds
 &\le\frac4\gamma\Bigl[F_{t,+}+F_{t,-}\\
 &\hspace{8mm}+\sqrt{m_tF_{t,+}}+\sqrt{m_tF_{t,-}}\Bigr]\\
 &\le\frac4\gamma\left[F_H+\sqrt{KF_H}\right].
\end{align*}
The case $b_t=0$ is immediate. The channel factorization and change of
time variable in Appendix~\ref{app:thermalmoment} give
\[
 \int_0^\infty\norm{T_sb_t}_1ds
 \le\int_0^\infty\norm{T_s^{(0)}b_t}_1ds.
\]
The phase rotations do not affect these norms. For
$x=\rho-\tau_{\bar n}$, the energy bound gives
$\norm{T_sx}_1\le C_Ee^{-\gamma s/2}$ on $\mathcal S_E$.
Substitution in \eqref{eq:cesaroestimate} proves
\eqref{eq:oscforcingbound}.
\end{proof}

\subsection{Reset formula and integrated response}\label{app:oscreset}
For a density operator $\sigma$, write $P_\sigma(X)=\tr(X)\sigma$
for the replacement channel.
Choose states $\sigma_t$, numbers $0\le a_t\le A_0<\infty$, and set
\begin{equation}\label{eq:oscresetD}
 D_t=a_t(P_{\sigma_t}-\id).
\end{equation}
These are bounded generators of CPTP semigroups, with
$\norm{D_t}\le2A_0$. Thus the abstract domain construction applies and
every $\Phi_t$ is CPTP. The standard renewal representation for quantum
resetting~\cite{MukherjeeSenguptaMajumdar2018,RoseEtAl2018,PerfettoCarolloLesanovsky2022}
gives the exact map
\begin{equation}\label{eq:oscreset}
\begin{split}
 \Phi_t={}&e^{-a_t}T_t\\
 &+\frac{a_t}{t}\int_0^t e^{-a_ts/t}P_{T_s\sigma_t}\,ds.
\end{split}
\end{equation}
The coefficients form a convex mixture with total weight one.

For fixed $t$, the generator is
$L-a_t\id/t+a_tP_{\sigma_t}/t$ and the trace is conserved.
Solving its inhomogeneous equation on a trace-one input gives
\eqref{eq:oscreset}; linearity extends the formula to the trace class.

For moving targets, a uniform bound
$\sup_{t\ge t_0}\tr(N\sigma_t)\le S<\infty$ gives
\begin{equation}\label{eq:oscresetrate}
 \sup_{\rho\in\mathcal S_E}\norm{\Phi_t\rho-\tau_{\bar n}}_1
 \le C_E e^{-\gamma t/2}+\frac{2A_0C_S}{\gamma t},\qquad t\ge t_0.
\end{equation}
Indeed, subtract $\tau_{\bar n}$ in \eqref{eq:oscreset}, apply
Lemma~\ref{prop:oscenergy} to both inputs, and integrate
$C_S e^{-\gamma s/2}$ over $[0,\infty)$.
For a fixed target, the leading response and its domain properties
follow under the weaker logarithmic-moment assumption from
Theorem~\ref{thm:thermalmoment}.

The energy bound also makes the forcing set relatively compact:
with $V=N+I$, the weighted trace norm of
$a_t(\sigma_t-\tau_{\bar n})$ is at most $A_0(S+\bar n+2)$.
Appendix~\ref{app:compact} and Theorem~\ref{thm:strong} therefore give
relaxation for every fixed normal initial state, even one of infinite
mean energy. The explicit uniform rate above concerns $\mathcal S_E$.

\subsection{Exact inverse-time relaxation}\label{app:exactinverse}
\begin{proposition}[Exact inverse-time relaxation]\label{prop:exactinverse}
Let $L=-i\omega[N,\,\cdot\,]+\gamma\D[\hat a]$ and let $D_t$ be given by
\eqref{eq:switchon}. Then $\Phi_t=e^{tL+D_t}$, completed by $\Phi_0=\id$,
is a strongly continuous family of CPTP maps. For every state $\rho$ and
$t>0$,
\begin{equation}\label{eq:exactonephoton}
 \Phi_t\rho-\pi_0
 =e^{-a_t}(T_t\rho-\pi_0)+p_t(\pi_1-\pi_0),
\end{equation}
where $p_t$ is defined in \eqref{eq:onephotonp} and $p_0=0$ by continuity. In particular, the vacuum input evolves as
\begin{equation}\label{eq:onephotonstate}
 \Phi_t\pi_0=(1-p_t)\pi_0+p_t\pi_1,
\end{equation}
and its exact distance to the asymptotic state is
\begin{equation}\label{eq:onephoton}
 \norm{\Phi_t\pi_0-\pi_0}_1
 =\frac{2a_t}{\gamma t+a_t}(1-e^{-\gamma t-a_t})
 \sim\frac{2a}{\gamma t}.
\end{equation}
More generally, for every finite $E$,
\begin{align}\label{eq:onephotonuniform}
 \sup_{\rho\in\mathcal S_E}
 \norm{t(\Phi_t\rho-\pi_0)-\frac a\gamma(\pi_1-\pi_0)}_1
 &\longrightarrow0,\\
 \sup_{\rho\in\mathcal S_E}
 \left|t\norm{\Phi_t\rho-\pi_0}_1-\frac{2a}\gamma\right|
 &\longrightarrow0.\label{eq:onephotoncoefficient}
\end{align}
Thus the relaxation rate is exactly $\Theta(t^{-1})$ for every initial
state with finite mean energy, with a strictly positive coefficient that
does not depend on that initial state.
\end{proposition}

\begin{proof}[Proof of Proposition~\ref{prop:exactinverse}]

Each frozen generator $L+D_t/t$ generates a CPTP semigroup. Formula
\eqref{eq:oscreset} gives strong continuity for $t>0$. At the origin,
$a_t\to0$ and $T_t\to\id$ strongly, while the integral term in that
formula has induced norm $1-e^{-a_t}\to0$.

Under pure loss, the one-photon state satisfies
$T_s\pi_1=\pi_0+e^{-\gamma s}(\pi_1-\pi_0)$.
Substitution in \eqref{eq:oscreset} yields
\[
 p_t=\frac{a_t}{t}\int_0^t
 e^{-(\gamma+a_t/t)s}ds,
\]
which proves \eqref{eq:exactonephoton} and \eqref{eq:onephotonp}.
The two projections are orthogonal, so
$\norm{\pi_1-\pi_0}_1=2$. This proves the exact vacuum distance.
Moreover, $tp_t\to a/\gamma$, while \eqref{eq:energybound} gives
\[
 \sup_{\rho\in\mathcal S_E}
 t e^{-a_t}\norm{T_t\rho-\pi_0}_1
 \le t\pi\sqrt E\,e^{-\gamma t/2}\longrightarrow0.
\]
The two uniform limits follow from \eqref{eq:exactonephoton} and the
reverse triangle inequality. Substituting $\rho=\pi_1$ also proves
\eqref{eq:excitederror}.
\end{proof}
\subsection{Number-state response}\label{app:energyresponse}
Under pure loss, the vacuum population of $T_s\pi_n$ is
$(1-e^{-\gamma s})^n$. All nonvacuum diagonal entries of
$T_s\pi_n-\pi_0$ are nonnegative and its vacuum entry is nonpositive.
Thus, for $n\ge1$,
\begin{align*}
 \norm{h_{\pi_n}}_1
 &=2\int_0^\infty[1-(1-e^{-\gamma s})^n]ds\\
 &=\frac2\gamma\int_0^1\frac{1-(1-u)^n}{u}\,du
 =\frac{2H_n}{\gamma}.
\end{align*}
Here $u=e^{-\gamma s}$ and
$[1-(1-u)^n]/u=\sum_{j=0}^{n-1}(1-u)^j$.
The case $n=0$ is immediate. This proves \eqref{eq:harmonicresponse}.

\subsection{Diagonal responses and infinite-energy resets}\label{app:logmoment}
Theorem~\ref{thm:thermalmoment} includes pure loss. In this subsection
write $h_\sigma=h_\sigma^{(0)}$ and assume $\mathcal M_\sigma<\infty$.
For a diagonal target $\sigma=\sum_n p_n\pi_n$, the response is explicit:
\begin{equation}\label{eq:diagonalresponse}
 h_\sigma=\frac1\gamma\left[
 \sum_{j\ge1}\frac{\sum_{n\ge j}p_n}{j}\pi_j
 -\mathcal M_\sigma\pi_0\right].
\end{equation}
To derive it, use the binomial pure-loss distribution. For $1\le j\le n$,
\[
 \int_0^\infty\tr(\pi_jT_s\pi_n)\,ds
 =\frac1\gamma\binom nj\int_0^1u^{j-1}(1-u)^{n-j}du
 =\frac1{\gamma j}.
\]
Tonelli then gives \eqref{eq:diagonalresponse}, with
$\norm{h_\sigma}_1=2\mathcal M_\sigma/\gamma$.
For a nondiagonal target the diagonal entries are unchanged by
number-basis coherences, so a vacuum measurement still gives
$\norm{h_\sigma}_1\ge2\mathcal M_\sigma/\gamma$.

For example, $p_n=[n(n+1)]^{-1}$ for $n\ge1$ has tail
$\sum_{n\ge j}p_n=1/j$. Its mean energy is infinite, whereas
$\mathcal M_\sigma=\sum_{j\ge1}j^{-2}=\pi^2/6$. Hence
\begin{equation}\label{eq:infiniteenergyinverse}
 \norm{\Phi_t\rho-\pi_0}_1\sim\frac{a\pi^2}{3\gamma t},
\end{equation}
uniformly on each $\mathcal S_E$. This illustrates why finite mean
energy is stronger than necessary for the reset target.

\subsection{Thermal logarithmic-moment criterion}\label{app:thermalmoment}
We prove the unified criterion together with useful quantitative
and pointwise refinements. If $\mathcal M_\sigma<\infty$, then
\begin{align}
 I_{\bar n}(\sigma)&\ge
 \frac{2|\mathcal M_\sigma-\log(1+\bar n)|}{\gamma(1+\bar n)},
 \label{eq:thermalIlower}\\
 I_{\bar n}(\sigma)&\le
 \frac4\gamma(\mathcal M_\sigma+\sqrt{\mathcal M_\sigma})
 +\frac{2\bar n}{\gamma(1+\bar n)}.
 \label{eq:thermalIupper}
\end{align}
The limits in Theorem~\ref{thm:thermalmoment} also hold pointwise for
each fixed initial state $\rho$ with $\mathcal M_\rho<\infty$.
Uniformity over classes with merely bounded logarithmic moment fails,
as the example at the end of this subsection shows. If $\mathcal M_\sigma=\infty$, the stronger conclusion is
\begin{equation}\label{eq:thermalinfinite}
 \inf_{\rho\text{ a state}}t\norm{\Phi_t\rho-\tau_{\bar n}}_1
 \longrightarrow\infty.
\end{equation}
The vacuum identity \eqref{eq:thermalvacidentity} holds in the extended
sense, including when $\mathcal M_\sigma=\infty$; its proof below also
establishes integrability of the negative part.
All statements concern a fixed finite $\bar n$.

\begin{proof}[Proof of Theorem~\ref{thm:thermalmoment} and the supplementary estimates]
Set $b=\bar n$, $B=1+b$, $\eta_s=e^{-\gamma s}$,
$g_s=1+b(1-\eta_s)$, and $z_s=\eta_s/g_s$.
Let $\mathcal L_z$ denote pure loss with transmissivity $z$ and
$\mathcal A_g$ a quantum-limited amplifier of gain $g$.
The standard Gaussian-channel decomposition~\cite{GarciaPatronEtAl2012} is
\begin{equation}\label{eq:thermalfactorization}
 T_s=\mathcal U_{\omega s}\mathcal A_{g_s}\mathcal L_{z_s},
 \qquad \mathcal A_g\pi_0=\tau_{g-1},
\end{equation}
where $\mathcal U_\theta(X)=e^{-i\theta N}Xe^{i\theta N}$.
All factors are CPTP. Writing $r=-\gamma^{-1}\log z_s$, one has
$r(0)=0$, $r(s)\to\infty$, and
\begin{equation}\label{eq:thermaltimechange}
 \frac{dr}{ds}=1+bz_s\ge1.
\end{equation}
For $0\le m\le b$, comparison of the geometric populations gives
$\tau_b\ge(1+m)\tau_m/B$. Subtracting this positive multiple and
using the triangle inequality therefore gives
\[
 \norm{\tau_b-\tau_m}_1\le\frac{2(b-m)}B.
\]
Contractivity in \eqref{eq:thermalfactorization} yields
\[
 \norm{T_s\sigma-\tau_b}_1
 \le\norm{\mathcal L_{z_s}\sigma-\pi_0}_1+\frac{2b\eta_s}B.
\]
Integrating, using \eqref{eq:thermaltimechange} and
Lemma~\ref{lem:logintegrability}, proves \eqref{eq:thermalIupper}.
The same contractivity and change of variable show, for any trace-class $Y$,
\begin{equation}\label{eq:thermalsignedintegral}
 \int_0^\infty\norm{T_sY}_1ds
 \le\int_0^\infty\norm{T_s^{(0)}Y}_1ds,
\end{equation}
where rotations on either side leave the trace norms unchanged.

For necessity, use the amplifier matrix elements
\[
 \langle0|\mathcal A_g(|m\rangle\langle n|)|0\rangle
 =\frac{\delta_{m0}\delta_{n0}}g.
\]
The quantum-limited amplifier only adds quanta; vacuum input remains
vacuum with probability $1/g$. Phase covariance removes off-diagonal
contributions to this population. Thus
$\mathcal A_g^*(\pi_0)=\pi_0/g$, and pure loss gives
\begin{equation}\label{eq:thermalvacprob}
 \tr(\pi_0T_s\sigma)
 =\frac1{g_s}\sum_{n\ge0}p_n(1-z_s)^n,
 \qquad p_n=\langle n|\sigma|n\rangle.
\end{equation}
Thus, for an arbitrary state including one with coherences,
\begin{align}
 \delta_\sigma(s)&:=\tr[\pi_0(\tau_b-T_s\sigma)]
                  =F_\sigma(s)-e_b(s),\label{eq:thermaldeficit}\\
 F_\sigma(s)&=\frac1{g_s}\sum_{n\ge0}p_n[1-(1-z_s)^n]\ge0,\notag\\
 e_b(s)&=\frac1{g_s}-\frac1B=\frac{b\eta_s}{Bg_s}\ge0.\notag
\end{align}
The substitution $z=\eta/[B-b\eta]$ gives
\[
 \eta=\frac{Bz}{1+bz},\qquad
 \frac{d\eta}{\eta}=\frac{dz}{z(1+bz)},\qquad
 \frac1g=\frac{1+bz}B.
\]
Tonelli and the harmonic-number integral imply
\begin{equation}\label{eq:thermaldeficitintegrals}
 \int_0^\infty F_\sigma(s)ds=\frac{\mathcal M_\sigma}{\gamma B},
 \qquad
 \int_0^\infty e_b(s)ds=\frac{\log B}{\gamma B}.
\end{equation}
This proves \eqref{eq:thermalvacidentity}, including its extended-value
interpretation. A vacuum measurement gives
$\norm{T_s\sigma-\tau_b}_1\ge2|\delta_\sigma(s)|$.
For finite $\mathcal M_\sigma$, integration proves
\eqref{eq:thermalIlower}. If $\mathcal M_\sigma=\infty$, the positive
term in \eqref{eq:thermaldeficit} has infinite integral and the negative
term has finite integral. Hence $I_b(\sigma)=\infty$.

For the reset family, the exact decomposition is
\begin{equation}\label{eq:thermalresetdecomposition}
\begin{split}
 \Phi_t\rho-\tau_b={}&e^{-a_t}(T_t\rho-\tau_b)\\
 &+\frac{a_t}{t}\int_0^t e^{-a_ts/t}(T_s\sigma-\tau_b)ds.
\end{split}
\end{equation}
Strong relaxation and Ces\`aro averaging make the integral term tend
to zero for every $\sigma$; the transient vanishes uniformly on
$\mathcal S_E$. If $\mathcal M_\sigma<\infty$, dominated convergence
after multiplication by $t$ gives \eqref{eq:thermalresetlimit}.
The finite-integral generator identity and closedness of $L$ give
$h_\sigma^{(b)}\in\dom L$ and $Lh_\sigma^{(b)}=\tau_b-\sigma$.
In particular, $h_\sigma^{(b)}=0$ implies $\sigma=\tau_b$.
For a fixed initial state with finite logarithmic moment,
$f(t)=\norm{T_t\rho-\tau_b}_1$ is decreasing and integrable, so
$tf(t)\le2\int_{t/2}^tf(s)ds\to0$. This gives the pointwise extension.

Finally, $\delta_\rho(t)\ge-e_b(t)$ for every initial state.
Applying the vacuum measurement to \eqref{eq:thermalresetdecomposition}
and using $e^{-a_ts/t}\ge e^{-A_0}$ gives
\begin{equation}\label{eq:thermalresetlower}
\begin{split}
 t\norm{\Phi_t\rho-\tau_b}_1
 \ge{}&2a_t e^{-A_0}\int_0^tF_\sigma(s)ds\\
 &-\frac{2a_t\log B}{\gamma B}-2t e^{-a_t}e_b(t).
\end{split}
\end{equation}
Since $e_b(t)\le b e^{-\gamma t}/B$, the last term vanishes
uniformly over all initial states. If $\mathcal M_\sigma=\infty$,
the first term diverges, proving \eqref{eq:thermalinfinite} and
the necessity in \eqref{eq:thermalresetiff}.
\end{proof}

The distinction between pointwise and uniform control is already visible
at zero temperature. Fix $m>0$, $c>1$, put
$n_t=\lceil e^{c\gamma t}\rceil$, and, for sufficiently large $t$, set
\[
 \rho_t=\left(1-\frac m{H_{n_t}}\right)\pi_0
       +\frac m{H_{n_t}}\pi_{n_t}.
\]
Then $\mathcal M_{\rho_t}=m$, but
\begin{equation}\label{eq:logmomentnonuniform}
\begin{split}
 t\norm{T_t\rho_t-\pi_0}_1
 &=\frac{2mt}{H_{n_t}}\left[1-(1-e^{-\gamma t})^{n_t}\right]\\
 &\longrightarrow\frac{2m}{c\gamma}>0.
\end{split}
\end{equation}
Here $H_{n_t}\sim c\gamma t$ and $n_te^{-\gamma t}\to\infty$.
Thus a bounded logarithmic moment alone does not give a uniform
$o(t^{-1})$ transient, although each fixed state with that moment
has this decay.

\subsection{Vacuum resetting in a thermal bath}\label{app:thermalvacuum}
An explicit finite-temperature example is resetting to the vacuum.
For $\bar n>0$, its response is
\begin{equation}\label{eq:thermalvacresponse}
 h_{\pi_0}^{(\bar n)}
 =\frac1\gamma\tau_{\bar n}\bigl[\log(1+\bar n)I-H_N\bigr].
\end{equation}
Consequently, uniformly on $\mathcal S_E$,
\begin{equation}\label{eq:thermalvacrate}
 \norm{\Phi_t\rho-\tau_{\bar n}}_1
 \sim\frac{a}{\gamma t}
 \tr\!\left[\tau_{\bar n}|H_N-\log(1+\bar n)I|\right].
\end{equation}
The coefficient is strictly positive. For $0<\bar n\le e-1$ it reduces
to $2a\log(1+\bar n)/[\gamma(1+\bar n)]$. In particular, at $\bar n=1$,
\begin{equation}\label{eq:thermaloneexample}
 \norm{\Phi_t\rho-\tau_1}_1\sim\frac{a\log2}{\gamma t}.
\end{equation}
The bath repopulates excited
levels while the weak reset favors the vacuum, leaving a displacement
of order $1/t$.

For $b=\bar n>0$, put $q=b/(1+b)$ and
$w_n=(1-q)q^n$, the thermal populations. The response already exists
in $\dom L$ by Theorem~\ref{thm:thermalmoment}. Since
$T_s\pi_0=\tau_{b(1-e^{-\gamma s})}$, put
$A_z=1+b(1-z)$. The population generating function, for $0\le z<1$, is
\begin{align*}
 \tr(z^N h_{\pi_0}^{(b)})
 &=\frac1\gamma\int_0^1
 \left[\frac1{A_z-(A_z-1)u}-\frac1{A_z}\right]\frac{du}{u}\\
 &=\frac{\log A_z}{\gamma A_z}.
\end{align*}
The identity $\sum_{n\ge0}H_nx^n=-\log(1-x)/(1-x)$ identifies
the coefficients as $w_n[\log(1+b)-H_n]/\gamma$, proving
\eqref{eq:thermalvacresponse}. Their trace is zero because
\[
 \sum_{n\ge0}w_nH_n=\sum_{j\ge1}\frac{q^j}{j}=\log(1+b).
\]
Taking the trace norm gives \eqref{eq:thermalvacrate}. For
$0<b\le e-1$, $\log(1+b)\le H_1=1$, so only the vacuum
coefficient is positive. The norm is then
$2\log(1+b)/[\gamma(1+b)]$, including \eqref{eq:thermaloneexample}.

\subsection{Two-level reset with a coherence}\label{app:coherentreset}
For the following explicit calculation we return to pure loss, $\bar n=0$.
Consider a reset supported on the vacuum and one-photon states,
\begin{equation}\label{eq:coherentreset}
 \sigma=(1-p)\pi_0+p\pi_1
 +c|1\rangle\langle0|+\bar c|0\rangle\langle1|,
\end{equation}
where $0\le p\le1$ and $|c|^2\le p(1-p)$. Its response is
\begin{equation}\label{eq:coherentresponse}
 h_\sigma=\frac p\gamma(\pi_1-\pi_0)
 +\frac{c}{\gamma/2+i\omega}|1\rangle\langle0|
 +\frac{\bar c}{\gamma/2-i\omega}|0\rangle\langle1|.
\end{equation}
Consequently the inverse-time coefficient is
\begin{equation}\label{eq:coherentcoefficient}
 a\norm{h_\sigma}_1
 =2a\sqrt{\frac{p^2}{\gamma^2}
              +\frac{|c|^2}{(\gamma/2)^2+\omega^2}}.
\end{equation}
The population moment is $\mathcal M_\sigma=p$ for all $c$, but the
coherence contributes to the response and is reduced by phase rotation.
For $p=c=1/2$ and $\omega=0$, the coefficient is $a\sqrt5/\gamma$;
dephasing the reset state changes it to $a/\gamma$.
Thus a population calculation alone gives the threshold, but generally
not the trace-norm coefficient.

For \eqref{eq:coherentreset}, pure loss and phase rotation give
\[
 T_s\sigma-\pi_0=p e^{-\gamma s}(\pi_1-\pi_0)
 +c e^{-(\gamma/2+i\omega)s}|1\rangle\langle0|+\mathrm{h.c.}
\]
Integration yields \eqref{eq:coherentresponse}. Its eigenvalues on
$\operatorname{span}\{|0\rangle,|1\rangle\}$ are
$\pm\sqrt{p^2/\gamma^2+|c|^2/[(\gamma/2)^2+\omega^2]}$,
which proves \eqref{eq:coherentcoefficient}.
The positivity constraint $|c|^2\le p(1-p)$ follows from the determinant
of the reset density matrix. No oscillator truncation is involved:
pure loss preserves this two-dimensional support.

\subsection{Increasing reset energy}\label{app:increasing}
Consider zero temperature and a constant reset strength $a>0$, but let
$\sigma_t=\pi_{n_t}$, where $n_t=\lceil e^{2\gamma t}\rceil$.
The perturbations remain uniformly bounded. Under pure loss, the vacuum
population of $T_s\pi_{n_t}$ is $(1-e^{-\gamma s})^{n_t}$. Uniformly for
$0\le s\le t$,
\[
 (1-e^{-\gamma s})^{n_t}
 \le e^{-n_te^{-\gamma t}}\longrightarrow0.
\]
Formula \eqref{eq:oscreset}, applied to the vacuum, therefore implies
\begin{equation}\label{eq:oscfailure}
 \norm{\Phi_t\pi_0-\pi_0}_1\longrightarrow2(1-e^{-a})>0.
\end{equation}
All states in this calculation are diagonal, so their distance to the
vacuum equals twice their nonvacuum population. This example retains
complete positivity and bounded perturbation norm; what fails is control
of the energy injected by the perturbation.

\subsection{Heavy-tail reset state}\label{app:logarithmic}
\begin{observation}[A logarithmic correction]\label{obs:logarithmic}
For the pure-loss oscillator and the reset state in
\eqref{eq:logtailmain}, the fixed bounded perturbation
$D=a(P_\sigma-\id)$, $a>0$, preserves strong relaxation, but
the asymptotic law \eqref{eq:logratemain} holds. Its logarithmic
moment is infinite.
\end{observation}

The compact-forcing theorem gives convergence. To determine its rate,
put $c=6/\pi^2$. The nonvacuum population at time $s$ is
\begin{equation}\label{eq:qtail}
 q(s)=\sum_{k\ge1}\frac{c}{k^2}
 \left[1-(1-e^{-\gamma s})^{\lceil e^k\rceil}\right].
\end{equation}
We claim $q(s)\sim c/(\gamma s)$. Set $u=\gamma s$ and fix $0<\epsilon<1$.
Since
\[
 1-e^{-e^{k-u}}
 \le1-(1-e^{-u})^{\lceil e^k\rceil}
 \le\min\{1,2e^{k-u}\},
\]
splitting the sum at $(1-\epsilon)u$ and $(1+\epsilon)u$ gives
\begin{align*}
 q(u/\gamma)&\le2e^{-\epsilon u}
 +c\sum_{k>\lfloor(1-\epsilon)u\rfloor}k^{-2},\\
 q(u/\gamma)&\ge(1-e^{-e^{\epsilon u}})
 c\sum_{k\ge\lceil(1+\epsilon)u\rceil}k^{-2}.
\end{align*}
Using $\sum_{k>m}k^{-2}\sim1/m$ and then sending $\epsilon\downarrow0$
proves the claim. In particular,
$\int_0^tq(s)ds\sim(c/\gamma)\log(1+\gamma t)$; no singularity occurs
at the origin, where $q(0)=1$ and $q(s)\le1$.

For the vacuum initial state, \eqref{eq:oscreset} gives exactly
\begin{equation}\label{eq:lograte}
 \norm{\Phi_t\pi_0-\pi_0}_1
 =\frac{2a}{t}\int_0^t e^{-as/t}q(s)\,ds
 \sim\frac{12a}{\pi^2\gamma}\frac{\log(1+\gamma t)}{t}.
\end{equation}
To justify the last step, the established asymptotic implies
$M_q=\sup_{s\ge0}s q(s)<\infty$. Therefore
\[
 0\le\int_0^t(1-e^{-as/t})q(s)ds
 \le\frac at\int_0^t s q(s)ds\le aM_q.
\]
The exponential weight changes the integral only by a bounded amount.
Thus a fixed bounded CPTP reset perturbation may preserve the limiting
state while violating an inverse-time upper bound, even for the vacuum
initial state. The obstruction is the nonintegrable relaxation tail of
the reset state.

The same asymptotic is compatible with continuous switch-on. Put
$R_\sigma=P_\sigma-\id$ and $a_t=a(1-e^{-\gamma t})$.
Both frozen generators in the following Duhamel estimate generate
CPTP contraction semigroups, so
\begin{equation}
\begin{split}
 \norm{e^{tL+a_tR_\sigma}-e^{tL+aR_\sigma}}_{1\to1}
 &\le |a_t-a|\norm{R_\sigma}\\
 &\le 2a e^{-\gamma t}.
\end{split}
\end{equation}
This exponentially small change preserves \eqref{eq:logratemain}.
Since $a_t\to0$ at the origin, the renewal formula also gives strong
continuity there, with $\Phi_0=\id$.

For the finite cutoff $\sigma_K$ defined in Section~\ref{sec:logtail},
the estimate $H_{\lceil e^k\rceil}=k+O(1)$ gives
\[
 \mathcal M_{\sigma_K}
 =c_K\sum_{k=1}^K\frac{H_{\lceil e^k\rceil}}{k^2}
 =c_K\log K+O(1).
\]
For each fixed $K$, the diagonal response
\eqref{eq:diagonalresponse} has norm $2\mathcal M_{\sigma_K}/\gamma$.
Theorem~\ref{thm:thermalmoment} therefore proves \eqref{eq:cutoffrate},
uniformly on each $\mathcal S_E$ at fixed $K$. This statement does not
give a remainder uniform in $K$, and hence does not exchange the
large-time and large-cutoff limits.

\subsection{Bounds from a truncated logarithmic moment}\label{app:truncatedmoment}
\begin{proof}[Proof of Proposition~\ref{prop:truncatedmoment}]
We consider pure loss, $D=a(P_\sigma-\id)$ with $a>0$, and vacuum
input. The phase rotation may be arbitrary. Put $x=\gamma t$,
$p_n=\langle n|\sigma|n\rangle$, and
\[
 f_n(u)=1-(1-e^{-u})^n,\qquad J_n(x)=\int_0^x f_n(u)\,du.
\]
The harmonic-number integral gives $J_n(\infty)=H_n$, while
$0\le f_n\le1$. Hence $J_n(x)\le\min(H_n,x)$.
For $n\ge1$ and $0\le u\le H_n$, the inequality
$H_n\le1+\log n$ gives $ne^{-u}\ge e^{-1}$, and therefore
\[
 f_n(u)\ge1-\exp(-ne^{-u})\ge1-\exp(-e^{-1})=c_0.
\]
The case $n=0$ is immediate. Consequently,
\begin{equation}\label{eq:truncatedJn}
 c_0\min(H_n,x)\le J_n(x)\le\min(H_n,x).
\end{equation}
Let $q(s)=\tr[(I-\pi_0)T_s\sigma]$ and
$Q(t)=\int_0^tq(s)\,ds$. Tonelli's theorem and
\eqref{eq:truncatedJn} yield
\begin{equation}\label{eq:truncatedQ}
 c_0m_\sigma(x)\le\gamma Q(t)\le m_\sigma(x).
\end{equation}
These quantities are finite at every finite $t$, even when
$\mathcal M_\sigma=\infty$.

For vacuum input, the renewal formula and a vacuum measurement give
\begin{equation}\label{eq:truncatedvacuumlower}
 \norm{\Phi_t\pi_0-\pi_0}_1
 \ge\frac{2a}{t}\int_0^t e^{-as/t}q(s)\,ds
 \ge\frac{2ae^{-a}}t Q(t).
\end{equation}
If $\sigma$ is diagonal, the first inequality is an equality: every
nonvacuum diagonal entry of the integrated response is nonnegative,
and its vacuum entry is nonpositive. The bounds
$e^{-a}\le e^{-as/t}\le1$ and \eqref{eq:truncatedQ} then prove
\eqref{eq:truncateddiagonal}.

For arbitrary coherences, integrate \eqref{eq:coherencehalftime} only
up to $t$. Cauchy--Schwarz gives
\begin{align}
 \int_0^t\norm{T_s\sigma-\pi_0}_1\,ds
 &\le4Q(t/2)
 +4\sqrt{\frac{(1-e^{-\gamma t/2})Q(t/2)}\gamma}\notag\\
 &\le\frac4\gamma\left[m_\sigma(x)+\sqrt{m_\sigma(x)}\right].
 \label{eq:truncatedcoherenceintegral}
\end{align}
Applying the triangle inequality to the renewal integral and using
$e^{-as/t}\le1$ proves the upper bound in
\eqref{eq:truncatedcoherent}; \eqref{eq:truncatedvacuumlower} proves
its lower bound. If $\mathcal M_\sigma=\infty$, then
$m_\sigma(x)\to\infty$, so the square-root term is eventually bounded
by $m_\sigma(x)$. This proves the asserted comparison of rates.
\end{proof}

The square-root term cannot be removed with a constant uniform over
all reset states with coherences. At $\omega=0$, take
$\sigma_p=|\psi_p\rangle\langle\psi_p|$, where
$|\psi_p\rangle=\sqrt{1-p}|0\rangle+\sqrt p|1\rangle$ and $0<p<1$.
Then $m_{\sigma_p}(x)\to p$, whereas
\eqref{eq:coherentcoefficient} gives
\[
 \lim_{t\to\infty}t\norm{\Phi_t\pi_0-\pi_0}_1
 =\frac{2a}{\gamma}\sqrt{4p-3p^2}.
\]
Its ratio to $p$ diverges as $p\downarrow0$.
For inputs in $\mathcal S_E$, the difference from the vacuum-input
output is at most $e^{-a}C_Ee^{-\gamma t/2}$, by
$\Phi_t(\rho-\pi_0)=e^{-a}T_t(\rho-\pi_0)$.
Thus the same estimates apply after adding this transient to the upper
bound and subtracting it from the lower bound.

\subsubsection{Finite-temperature bounds}\label{app:thermaltruncated}
Let $0\le b=\bar n<\infty$, $B=1+b$, $x=\gamma t$, and
$\Phi_t=e^{tL+a(P_\sigma-\id)}$, with $a>0$ fixed and $L$ the thermal
attenuation generator. For vacuum input, put
$d_{b,\sigma}(t)=\norm{\Phi_t\pi_0-\tau_b}_1$ and
\begin{equation}\label{eq:thermaltruncatedepsilon}
 \varepsilon_b(t)=\frac{2ab}{Bx}+\frac{2b}{B}e^{-a-x}.
\end{equation}
For every reset state and $t>0$,
\begin{align}
 d_{b,\sigma}(t)
 &\ge\frac{2ae^{-a}c_0}{Bx}m_\sigma(x)-\varepsilon_b(t),\notag\\
 d_{b,\sigma}(t)
 &\le\frac{4a}{x}\left[m_\sigma(x)+\sqrt{m_\sigma(x)}\right]
       +\varepsilon_b(t).
 \label{eq:thermaltruncatedbounds}
\end{align}
For diagonal $\sigma$, the upper bound improves to
$2am_\sigma(x)/x+\varepsilon_b(t)$.
The constants are independent of $\sigma$. Since
$\varepsilon_b(t)=O(x^{-1})$, an infinite logarithmic moment gives
$d_{b,\sigma}(t)\asymp m_\sigma(\gamma t)/(\gamma t)$ at every fixed
finite $b$. This comparison does not identify the asymptotic coefficient.

\begin{proof}[Derivation of the thermal bounds]
Use $g_s=B-be^{-\gamma s}$, $z_s=e^{-\gamma s}/g_s$, and
$v_s=-\gamma^{-1}\log z_s\ge s$, as in
\eqref{eq:thermalfactorization}. Write
$q_0(s)=\tr[(I-\pi_0)T_s^{(0)}\sigma]$ and
$Q_0(t)=\int_0^tq_0(s)\,ds$ for the corresponding pure-loss semigroup.
The identities $dv_s/ds=B/g_s$ and
$F_\sigma(s)=q_0(v_s)/g_s$ imply
\[
 \int_0^tF_\sigma(s)\,ds
 =\frac{Q_0(v_t)}B
 \ge\frac{c_0m_\sigma(\gamma t)}{\gamma B}.
\]
For vacuum input, the vacuum deficit of the initial transient is
$-e^{-a}e_b(t)$. Applying the vacuum measurement to the renewal formula,
using $e^{-as/t}\ge e^{-a}$ on the nonnegative $F_\sigma$ term, and
using
\[
 \int_0^te_b(s)\,ds\le\frac{\log B}{\gamma B}
 \le\frac b{\gamma B},\qquad e_b(t)\le\frac bB e^{-\gamma t},
\]
proves the lower bound in \eqref{eq:thermaltruncatedbounds}.

For the upper bound, the factorization and thermal-state estimate in
Appendix~\ref{app:thermalmoment} give
\begin{align*}
 \norm{T_s\sigma-\tau_b}_1
 &\le\norm{T_{v_s}^{(0)}\sigma-\pi_0}_1+\frac{2b}{B}e^{-\gamma s}\\
 &\le\norm{T_s^{(0)}\sigma-\pi_0}_1+\frac{2b}{B}e^{-\gamma s}.
\end{align*}
The last step uses contractivity of pure loss and $v_s\ge s$; phase
rotations do not change these norms. Integrate this inequality in the
renewal formula and apply \eqref{eq:truncatedcoherenceintegral}.
The initial transient is bounded by $2be^{-a-\gamma t}/B$, which
completes the upper estimate. For diagonal $\sigma$, use instead
$\norm{T_s^{(0)}\sigma-\pi_0}_1=2q_0(s)$ and
$\gamma Q_0(t)\le m_\sigma(\gamma t)$.
\end{proof}

\subsection{Weak resetting and stationary response}\label{app:weakreset}
Fix a thermal attenuation semigroup with $0\le\bar n<\infty$ and
a reset state $\sigma$. For a fixed rate $r>0$, denote the reset
semigroup by $S_s^{(r)}=e^{s[L+r(P_\sigma-\id)]}$.
The resolvent formula for a contraction semigroup gives
\begin{equation}\label{eq:nessresolvent}
 \rho_r^{\mathrm{ss}}
 =r\int_0^\infty e^{-rs}T_s\sigma\,ds
 =r(r-L)^{-1}\sigma.
\end{equation}
The integral is a density operator and belongs to $\dom L$.
Moreover, $(L-r)\rho_r^{\mathrm{ss}}=-r\sigma$, so
$[L+r(P_\sigma-\id)]\rho_r^{\mathrm{ss}}=0$.
The renewal formula gives the exact identity
\begin{equation}\label{eq:nesssemigroupidentity}
 S_s^{(r)}\rho-\rho_r^{\mathrm{ss}}
 =e^{-rs}T_s(\rho-\rho_r^{\mathrm{ss}}).
\end{equation}
It proves uniqueness of the stationary state and convergence to it
in trace norm. For $a_t>0$, taking $r=a_t/t$ and $s=t$ gives the decomposition
of the finite-horizon family into its stationary displacement and
transient.

If $\mathcal M_\sigma<\infty$, Theorem~\ref{thm:thermalmoment}
and dominated convergence imply
\[
 \frac{\rho_r^{\mathrm{ss}}-\tau_{\bar n}}r
 =\int_0^\infty e^{-rs}(T_s\sigma-\tau_{\bar n})\,ds
 \longrightarrow h_\sigma^{(\bar n)}
 \qquad(r\downarrow0)
\]
in trace norm, proving \eqref{eq:nessresponse}.
Conversely, the functions $F_\sigma$ and $e_{\bar n}$ defined in
\eqref{eq:thermaldeficit} satisfy
\begin{equation}\label{eq:nessinfinite}
\begin{split}
 \frac{\norm{\rho_r^{\mathrm{ss}}-\tau_{\bar n}}_1}{r}
 \ge{}&2\int_0^\infty e^{-rs}F_\sigma(s)\,ds\\
 &-2\int_0^\infty e^{-rs}e_{\bar n}(s)\,ds.
\end{split}
\end{equation}
If $\mathcal M_\sigma=\infty$, monotone convergence makes the first
integral diverge as $r\downarrow0$, while the second remains bounded
by \eqref{eq:thermaldeficitintegrals}. Thus the norm divided by $r$
diverges. This establishes both necessity of the moment condition
for a finite trace-class derivative and the equivalence
\[
 \norm{\rho_r^{\mathrm{ss}}-\tau_{\bar n}}_1=O(r)
 \quad\Longleftrightarrow\quad\mathcal M_\sigma<\infty.
\]

The stationary response also gives the leading finite-horizon response
when the moment is finite. Put
$f_\sigma(t)=\norm{T_t\sigma-\tau_{\bar n}}_1$.
Contractivity makes $f_\sigma$ decreasing, and
Theorem~\ref{thm:thermalmoment} makes it integrable. Hence
$tf_\sigma(t)\le2\int_{t/2}^t f_\sigma(s)\,ds\to0$.
For any $r>0$,
\[
 \norm{T_t\rho_r^{\mathrm{ss}}-\tau_{\bar n}}_1
 \le r\int_0^\infty e^{-rs}f_\sigma(t+s)\,ds
 \le f_\sigma(t).
\]
With $r_t=a_t/t$ and $a_t\to a>0$, \eqref{eq:nesssemigroupidentity}
therefore yields
\begin{equation}\label{eq:nessuniformtransient}
\begin{split}
 \sup_{\rho\in\mathcal S_E}
 \norm{\Phi_t\rho-\rho_{r_t}^{\mathrm{ss}}}_1
 &\le e^{-a_t}\left(C_Ee^{-\gamma t/2}+f_\sigma(t)\right)\\
 &=o(t^{-1}).
\end{split}
\end{equation}
This identifies the inverse-time coefficients of the two families.
For arbitrary $\sigma$, the same bound without the rate still proves
that their difference tends to zero uniformly on $\mathcal S_E$;
it need not be negligible relative to their leading errors.

To see the latter distinction, consider pure loss and a diagonal reset
with nonvacuum population $q(s)\sim Cs^{-\beta}$, where
$C>0$ and $0<\beta<1$. For fixed $a>0$, the exact formulas give
\begin{align}
 \norm{\Phi_t\pi_0-\pi_0}_1
 &\sim2aC\,t^{-\beta}\int_0^1e^{-au}u^{-\beta}\,du,
 \label{eq:nessfinitehorizontail}\\
 \norm{\rho_{a/t}^{\mathrm{ss}}-\pi_0}_1
 &\sim2aC\,t^{-\beta}\int_0^\infty e^{-au}u^{-\beta}\,du.
 \label{eq:nessinfinitetail}
\end{align}
Indeed, $q(s)\le C'(1+s)^{-\beta}$ for some $C'<\infty$,
so after setting $s=tu$, dominated convergence applies with majorant
$C'e^{-au}u^{-\beta}$. The omitted interval $u>1$ has a strictly
positive integral. The corresponding transient is therefore of
order $t^{-\beta}$, rather than a lower-order correction.
Such reset states exist: with
$Z_\beta=\sum_{k\ge1}k^{-1-\beta}$, take
\[
 \sigma=Z_\beta^{-1}\sum_{k\ge1}k^{-1-\beta}\pi_{\lceil e^k\rceil}.
\]
Splitting the sum at $(1-\epsilon)\gamma s$ and
$(1+\epsilon)\gamma s$, as in Appendix~\ref{app:logarithmic}, gives
$q(s)\sim(\beta Z_\beta\gamma^\beta)^{-1}s^{-\beta}$.

\subsubsection{Truncated-moment bounds for the stationary state}\label{app:nesstruncated}
For pure loss, the truncated moment also controls the stationary
displacement at every reset rate. Put
$d_\sigma^{\mathrm{ss}}(r)=\norm{\rho_r^{\mathrm{ss}}-\pi_0}_1$,
$m=m_\sigma(\gamma/r)$, and
$c_0=1-\exp(-e^{-1})$. For every $r>0$ and every reset state,
including arbitrary number-basis coherences,
\begin{equation}\label{eq:nesstruncatedcoherent}
 \frac{2c_0}{e}\frac r\gamma m
 \le d_\sigma^{\mathrm{ss}}(r)
 \le\frac{4r}{\gamma}(m+\sqrt m).
\end{equation}
For a diagonal reset, the upper bound improves to
\begin{equation}\label{eq:nesstruncateddiagonal}
 d_\sigma^{\mathrm{ss}}(r)\le\frac{2r}{\gamma}m.
\end{equation}
The phase rotation is unrestricted. These estimates require no finite
moment of $\sigma$, since the truncated moment is finite at every
positive $r$. The square-root term is needed for a bound uniform over all
reset states with coherences, as the two-level example in Appendix~\ref{app:truncatedmoment} shows.

\begin{proof}[Derivation of the stationary bounds]
For any nonnegative decreasing function $f$ and $T=1/r$,
\begin{equation}\label{eq:laplacetruncation}
 e^{-1}\int_0^T f(s)\,ds
 \le\int_0^\infty e^{-rs}f(s)\,ds
 \le\int_0^T f(s)\,ds.
\end{equation}
The lower bound follows by restricting the integral to $[0,T]$.
For the upper bound, the omitted tail is at most
$f(T)e^{-1}/r$, whereas the weight removed from the first interval
is at least
\[
 f(T)\int_0^T(1-e^{-rs})\,ds=f(T)e^{-1}/r.
\]

Under pure loss, both
$q(s)=\tr[(I-\pi_0)T_s\sigma]$ and
$f_\sigma(s)=\norm{T_s\sigma-\pi_0}_1$ are decreasing;
the latter follows from contractivity and invariance of $\pi_0$.
The stationary renewal formula and a vacuum measurement give
\[
 d_\sigma^{\mathrm{ss}}(r)
 \ge2r\int_0^\infty e^{-rs}q(s)\,ds,
\]
with equality for diagonal $\sigma$.
Apply \eqref{eq:laplacetruncation} to $q$ and then use
\eqref{eq:truncatedQ} at $T=1/r$. This proves the lower bound in
\eqref{eq:nesstruncatedcoherent} and the diagonal upper bound.
For arbitrary $\sigma$, the triangle inequality instead gives
\[
 d_\sigma^{\mathrm{ss}}(r)
 \le r\int_0^\infty e^{-rs}f_\sigma(s)\,ds
 \le r\int_0^{1/r}f_\sigma(s)\,ds.
\]
Equation~\eqref{eq:truncatedcoherenceintegral} proves the remaining
upper bound. When $\mathcal M_\sigma=\infty$, monotone convergence
gives $m_\sigma(\gamma/r)\to\infty$, so the square-root correction
is eventually bounded by the moment itself.

\end{proof}

\subsection{Independent modes with a product reset}\label{app:multimode}
A finite collection of independent modes admits an immediate extension
of the integrability criterion for product reset states. Let
$T_s=\bigotimes_{j=1}^dT_s^{(j)}$, where each factor is a thermal
attenuation semigroup with damping rate $\gamma_j>0$, finite bath
occupation $\bar n_j$, and stationary state $\tau_j$.
Let $\sigma=\bigotimes_{j=1}^d\sigma_j$ and
$\tau=\bigotimes_{j=1}^d\tau_j$. Partial trace and the telescoping
identity for tensor products give, at every $s\ge0$,
\[
 \max_j\norm{T_s^{(j)}\sigma_j-\tau_j}_1
 \le\norm{T_s\sigma-\tau}_1
 \le\sum_{j=1}^d\norm{T_s^{(j)}\sigma_j-\tau_j}_1.
\]
By Theorem~\ref{thm:thermalmoment}, the middle expression is integrable
over $s\ge0$ if and only if every marginal has a finite logarithmic
number moment. With $N_{\mathrm{tot}}=\sum_{j=1}^dN_j$, the commuting
number operators satisfy
\[
 \log(I+N_j)\le\log(I+N_{\mathrm{tot}})
 \le\sum_{j=1}^d\log(I+N_j).
\]
Thus
\begin{equation}\label{eq:multimodeintegrability}
\begin{gathered}
 \int_0^\infty\norm{T_s\sigma-\tau}_1\,ds<\infty
 \\
 \Longleftrightarrow\quad
 \tr[\sigma\log(I+N_{\mathrm{tot}})]<\infty.
\end{gathered}
\end{equation}
This argument concerns finitely many independent modes and product
reset states. It makes no claim for correlated reset states or coupled
quasi-free evolutions.

\section{Complete positivity}\label{app:positivity}
\subsection{Faithful finite-dimensional limit}\label{app:cp}
Trace annihilation and Hermiticity preservation imply that $\Phi_t$ is
trace-preserving and Hermiticity-preserving for every $t$. Positivity remains a separate question. In finite dimension, the
Choi criterion~\cite{Choi1975} leads to the following conclusion.

\begin{theorem}[Faithful limiting state]\label{thm:cp}
For a relaxing finite-dimensional $L$ with $\rho_*>0$, every perturbation satisfying
\eqref{eq:structure} and \eqref{eq:sublinear} gives a CPTP map $\Phi_t$ for all
sufficiently large $t$. More generally, for a fixed relaxing $L$,
\begin{equation}\label{eq:iffcp}
\begin{gathered}
 \rho_*>0\\
 \Updownarrow\\
 \text{Every bounded family $D_t$ satisfying \eqref{eq:structure}}\\
 \text{has $e^{tL+D_t}$ CPTP for all sufficiently large $t$.}
\end{gathered}
\end{equation}
\end{theorem}

In particular, an ergodic finite-dimensional Davies generator
\cite{Davies1974} at finite inverse temperature has a full-rank Gibbs stationary state, so
Theorem~\ref{thm:cp} applies. The theorem does not assert CPTP at early times,
CP divisibility of $\Phi_t$, or the GKSL property of $L+D_t/t$.
Related eventual entanglement-breaking properties for divisible dynamics
are studied in Ref.~\cite{SzczygielskiChruscinski2024}; divisibility is
not assumed here.

A sufficient condition for complete positivity at every time is that
each frozen generator itself be GKSL. This condition can be tested through
its Kossakowski matrix, as described in Appendix~\ref{app:kossakowski}.
It is stronger than necessary: a Pauli-channel example in
Appendix~\ref{app:pauli} is eventually CPTP even though its frozen generator
is never GKSL.

\begin{proof}[Proof of Theorem~\ref{thm:cp}]

Use the unnormalized Choi matrix
$J(T)=\sum_{ij}|i\rangle\langle j|\otimes T(|i\rangle\langle j|)$.
Then $J(P)=I_d\otimes\rho_*$ and its smallest eigenvalue is
$\mu=\lambda_{\min}(\rho_*)>0$. Operator-norm convergence $\Phi_t\to P$
implies $J(\Phi_t)\to J(P)$. For large $t$, Weyl's bound gives
$J(\Phi_t)>0$, and the Choi criterion~\cite{Choi1975} implies complete positivity.
A conservative explicit sufficient threshold is
\begin{equation}\label{eq:cptime}
 d^2\left[2M e^{-\beta_t t}
 +\frac{M(1-e^{-\beta_t t})}{\beta_t}\frac{f_t}{t}\right]<\mu,
\end{equation}
because $\norm{J(T)}_\infty\le d^2\norm T_{1\to1}$ by the triangle inequality
on the matrix units.

Conversely, suppose $\rho_*$ is singular. Choose a unit vector
$u\in\ker\rho_*$ and put $Z=\rho_*-|u\rangle\langle u|$. The constant map
\[
 D(X)=-\tr(X)L(Z)
\]
is bounded, Hermiticity-preserving and trace-annihilating, with $D|_\E=0$.
Formula \eqref{eq:variation} gives exactly
\[
 \Phi_t\rho_* =\rho_*+\frac{(I-e^{tL})Z}{t}.
\]
Its $u,u$ matrix element is
$[-1-\langle u|e^{tL}Z|u\rangle]/t<0$ for large $t$, since $e^{tL}Z\to0$.
Thus it is eventually nonpositive. This disproves the universal property
when $\rho_*$ is singular.
\end{proof}
\subsection{Kossakowski matrix conditions}\label{app:kossakowski}
In a fixed canonical basis of traceless operators, write the Kossakowski
matrices of $L$ and the dissipative part of $D_t$ as $c\ge0$ and
$\Delta_t=\Delta_t^\dagger$. The frozen operator $L+D_t/t$ is GKSL if and
only if $c+\Delta_t/t\ge0$. This is sufficient for CPTP of $\Phi_t$ but
is stronger than necessary.

If $c\ge\kappa I>0$ and $\norm{\Delta_t}\le R$, the frozen
generator is GKSL for $t\ge R/\kappa$. For singular $c$, boundedness
alone does not ensure this: the condition must also be checked on
$\ker c$ and in the off-diagonal blocks connecting it to
$\operatorname{supp}c$. Eventual complete positivity of the exponential
can nevertheless hold, as Appendix~\ref{app:pauli} demonstrates.

\subsection{Bounded perturbation with no eventual positivity}
Let $P_j(X)=\tr(X)|j\rangle\langle j|$ for a qubit and take
\[
 L=P_0-\id,\qquad D=P_0-P_1.
\]
The unperturbed dynamics relaxes to $|0\rangle\langle0|$. Nevertheless,
\begin{equation}
 e^{tL+D}X=e^{-t}X+(1-e^{-t})P_0X
 +\frac{1-e^{-t}}t(P_0-P_1)X,
\end{equation}
so the output of $|0\rangle\langle0|$ has population
$-(1-e^{-t})/t$ in $|1\rangle$. It is nonpositive for every $t>0$, despite
converging to $P_0$. A bounded continuous switch-on factor multiplying $D$,
equal to zero at zero and to one for $t\ge1$, gives the same eventual failure.

An even stronger distinction is possible: let
$L=\D[|0\rangle\langle1|]$ and
$D(X)=\kappa(X-\sigma_zX\sigma_z)/2$, $\kappa>0$, using
the convention \eqref{eq:dissipator}.
Here $D\rho_*=0$, so the map converges exponentially and has the same fixed
state. But the excited-state survival is $e^{-t}$ and the coherence
multiplier is $e^{-t/2+\kappa}$. Choi positivity requires
$e^{-t+2\kappa}\le e^{-t}$, which fails for every finite $t$.

\subsection{Eventual CPTP without a GKSL frozen generator}\label{app:pauli}
With Pauli matrices, take
\[
 L=\gamma\bigl(\D[\sigma_x]+\D[\sigma_y]\bigr),\qquad
 D=-a\D[\sigma_z],\qquad\gamma,a>0.
\]
The limiting state is $I/2$. In the Pauli representation the Kossakowski
coefficients of $L+D/t$ have signs $(+,+,-)$, so it is not GKSL for any finite
$t>0$. The exponential has Bloch multipliers
\[
 (u,u,w)=(e^{2a-2\gamma t},e^{2a-2\gamma t},e^{-4\gamma t}).
\]
Its unnormalized Choi eigenvalues are
$(1+2u+w)/2,(1-w)/2,(1-w)/2,(1-2u+w)/2$. Hence
\begin{equation}\label{eq:paulithreshold}
 e^{tL+D}\text{ is CPTP}\quad\Longleftrightarrow\quad
 t\ge\frac{\operatorname{arcosh}(e^{2a})}{2\gamma}.
\end{equation}
This also demonstrates why proving $L+D_t/t$ to be GKSL is an unnecessarily
strong route to eventual CPTP in the faithful case.

\section{Sharpness constructions}\label{app:sharpness}

\subsection{Exact finite-dimensional reset family}\label{app:finiteexamples}\label{sec:rates}
For a state $\varrho$, put $R_\varrho=P_\varrho-\id$. This is GKSL: if
$\varrho=\sum_i p_i|i\rangle\langle i|$, jumps
$\sqrt{p_i}|i\rangle\langle j|$ yield $R_\varrho$.
Choose distinct states $\rho_*,\sigma$ and set
$L=R_{\rho_*}$, $D_t=a(t)R_\sigma$, with $a(t)\ge0$.
For $t>0$, write $q_t=e^{-t-a(t)}$ and
$\varrho_t=(t\rho_*+a(t)\sigma)/(t+a(t))$. Direct exponentiation gives
\begin{equation}\label{eq:reset}
\begin{aligned}
 \Phi_t&=q_t\id+(1-q_t)P_{\varrho_t},\\
 \Phi_t\rho-\rho_*&=q_t(\rho-\rho_*)
 +(1-q_t)\frac{a(t)}{t+a(t)}(\sigma-\rho_*).
\end{aligned}
\end{equation}
The map is CPTP. For orthogonal pure $\rho_*,\sigma$,
\begin{equation}\label{eq:resetworst}
 W(t)=2\left[q_t+(1-q_t)\frac{a(t)}{t+a(t)}\right].
\end{equation}
The triangle inequality gives the upper bound and the input $\sigma$
attains it. The same expression is the induced $1\to1$ distance from
$P_{\rho_*}$. These formulas have two useful consequences.

First, bounded perturbations have no universal polynomial lower rate:
$a(t)=0$ gives $W(t)=2e^{-t}$, while constant $a(t)=a_0>0$ gives
$W(t)\sim2a_0/t$. Thus the action on equilibrium determines which
contribution survives. Second, $D_t=o(t)$ gives no universal polynomial
upper rate: $a(t)=t/\log(t+e)$ yields $W(t)\asymp1/\log t$.
At the linear scale $a(t)=\alpha t$, the limit becomes
$(\rho_*+\alpha\sigma)/(1+\alpha)$, so the original equilibrium
is no longer recovered. The large-time choices can be switched on
continuously with $a(0)=0$ without changing these conclusions.

\subsection{Defective generators}\label{app:defective}
The exponent slack in Theorem~\ref{thm:twosided} cannot generally be
removed, even for a bounded GKSL perturbation that leaves a faithful
stationary state fixed. Consider the three-level generator
\begin{equation}\label{eq:defectiveL}
 L=\D[|0\rangle\langle1|]+\D[|1\rangle\langle2|]
   +(\operatorname{diag}-\id)+R_{I/3},
\end{equation}
where $\operatorname{diag}$ is complete dephasing in the displayed basis.
Its stationary state is $\rho_*=\operatorname{diag}(7/12,1/4,1/6)$.
On diagonal trace-zero matrices, with coordinates $(x_1,x_2)$ and
$x_0=-x_1-x_2$, the restriction is
\[
 A_{\rm cl}=\begin{pmatrix}-2&1\\0&-2\end{pmatrix}.
\]
The coherence eigenvalues are $-5/2$ and $-3$, so the gap is $g=2$.
Now take the bounded GKSL perturbation
\begin{equation}\label{eq:defectiveD}
 D=\varepsilon\D[|2\rangle\langle1|]
   +\frac{3\varepsilon}{2}\D[|1\rangle\langle2|],
 \qquad\varepsilon>0.
\end{equation}
The two stationary fluxes agree,
$\varepsilon(1/4)=(3\varepsilon/2)(1/6)$, hence $D\rho_*=0$ and $f_t=0$.
On the population subspace,
\[
 tA_{\rm cl}+D_{\rm cl}
 =\begin{pmatrix}-2t-\varepsilon&t+3\varepsilon/2\\
                  \varepsilon&-2t-3\varepsilon/2\end{pmatrix},
\]
with eigenvalues
\begin{equation}\label{eq:defectivespectrum}
 \lambda_\pm(t)=-2t-\frac{5\varepsilon}{4}
     \pm\sqrt{\varepsilon t+\frac{25\varepsilon^2}{16}}.
\end{equation}
As in the proof of Theorem~\ref{thm:twosided}, the spectral radius on
the Hermitian trace-zero subspace bounds $W(t)$ from below. Therefore
\[
 \frac{W(t)}{e^{-2t}}
 \ge\exp\!\left[-\frac{5\varepsilon}{4}
          +\sqrt{\varepsilon t+\frac{25\varepsilon^2}{16}}\right]
 \longrightarrow\infty.
\]
No constant $C$ can give $W(t)\le C[f_t/t+e^{-gt}]$ here.
The example simultaneously exhibits unbounded absolute eigenvalue
shifts of order $\sqrt t$ under a fixed perturbation of the exponent,
while the normalized eigenvalues of $L+D/t$ still approach those of $L$.
This is consistent with the role of nonnormality in spectral convergence
estimates~\cite{SzehrReebWolf2015}.

\subsection{Infinite-dimensional counterexamples}\label{app:infiniteexamples}\label{sec:limitations}
The results of Sections~\ref{sec:strong} and \ref{sec:uniform} apply on an
arbitrary separable Hilbert space. However, several conclusions obtained
in Appendix~\ref{app:finite} use finite dimension essentially. We now give
examples that separate these conclusions, while retaining quantum Markov
dynamics for the unperturbed evolution.

\subsubsection{A bounded family of perturbations can destroy strong relaxation}
Let $\mathcal H=\ell^2(\mathbb N_0)$, write
$\pi_n=|n\rangle\langle n|$, and define the bounded positive operator
$F=\sum_{n\ge1}n^{-1}\pi_n$. The generator
\begin{equation}\label{eq:slowgenerator}
 L(X)=\tr(FX)\pi_0-\tfrac12\{F,X\}
\end{equation}
describes decay from level $n$ to the ground state at rate $1/n$.
Its CPTP semigroup is
\[
 T_sX=E_sXE_s+\tr[(I-E_s^2)X]\pi_0,\qquad E_s=e^{-sF/2}.
\]
Since $E_s\to\pi_0$ strongly, finite-rank approximation gives
$T_sX\to\tr(X)\pi_0$ in trace norm for every $X$. Nevertheless,
$\norm{T_s\pi_n-\pi_0}_1=2e^{-s/n}$, so convergence is not uniform over
states at any finite time.

For $t\ge1$, let $n(t)=\lceil t^2\rceil$ and
$D_t=\D[|n(t)\rangle\langle0|]$. These are bounded GKSL generators with
$\norm{D_t}\le2$. Starting from $\pi_0$, the map $e^{tL+D_t}$ has excited
population
\begin{equation}\label{eq:slowcounter}
 p_t=\frac{1-e^{-(1+t/n(t))}}{1+t/n(t)}\longrightarrow1-e^{-1}.
\end{equation}
This follows from the two-state rate equation with upward rate $1$ and
downward rate $t/n(t)$ in the exponent. Thus
$\norm{\Phi_t\pi_0-\pi_0}_1\to2(1-e^{-1})$, although every $\Phi_t$ is
CPTP. Here $D_t\pi_0=\pi_{n(t)}-\pi_0$ escapes every norm-compact set.
The example shows why boundedness alone does not replace
\eqref{eq:forcingcesaro} in Theorem~\ref{thm:strong}.

\subsubsection{Faithfulness does not imply eventual positivity}
On the same Hilbert space, set
\[
 \begin{gathered}
 \varrho=\sum_{n\ge0}\frac34\,4^{-n}\pi_n,\qquad
 Y=\pi_0-\sum_{n\ge1}2^{-n}\pi_n,\\
 L=P_\varrho-\id,\qquad D(X)=\tr(X)Y.
 \end{gathered}
\]
The stationary state $\varrho$ is faithful, $L$ relaxes uniformly
exponentially, and $D$ is bounded, Hermiticity-preserving and
trace-annihilating. Since $D|_\E=0$, the exact formula gives
\begin{equation}\label{eq:faithfulcounter}
 e^{tL+D}\varrho=\varrho+\frac{1-e^{-t}}tY.
\end{equation}
For every $t>0$, its $n$th diagonal entry is negative for sufficiently
large $n$, since $2^{-n}/4^{-n}\to\infty$. Hence the maps are never positive,
despite converging in induced trace norm to the faithful replacement map.
A faithful trace-class state on an infinite-dimensional space has no
strictly positive lower bound by a multiple of the identity. Accordingly,
the interior-point argument in Theorem~\ref{thm:cp} no longer applies.

A sufficient structural assumption is available in either dimension:
if each bounded $D_t$ itself generates a CPTP semigroup, then $L+D_t/t$
generates a CPTP semigroup. This follows from the product formula and the
bounded perturbation theorem. For more general $D_t$, positivity requires
an independent argument.

\subsubsection{A transient can disappear after finite time}
On $\mathcal H=\mathbb C|0\rangle\oplus L^2(0,1)$, set $V_s=1\oplus S_s$,
where $(S_sf)(x)=\mathbf1_{\{x\ge s\}}f(x-s)$ is the right shift
for $0\le s<1$, and $S_s=0$ for $s\ge1$. Define
\begin{equation}\label{eq:finiteextinction}
 T_sX=V_sXV_s^\dagger+
 \tr[(I-V_s^\dagger V_s)X]\pi_0.
\end{equation}
The maps are CPTP and strongly continuous by finite-rank approximation.
The semigroup identity follows from that of $V_s$ and
$I-V_{s+u}^\dagger V_{s+u}=(I-V_u^\dagger V_u)+
V_u^\dagger(I-V_s^\dagger V_s)V_u$.
For $s\ge1$, $T_s=P_{\pi_0}$ exactly. On $\E$ this semigroup
satisfies \eqref{eq:basebound} with $M=e^\gamma$ for any $\gamma>0$, yet
$W(t)=0$ for $t\ge1$ when $D_t=0$. Therefore a strictly positive
exponential lower bound on the transient cannot be inferred from
\eqref{eq:basebound}, whereas \eqref{eq:responseworst} remains valid.

\section{Cumulant exponents and time-local equations}\label{app:cumulant}
\subsection{A finite-dimensional correlation criterion}
Let $H_0$ act on a finite-dimensional system and fix a finite interaction
$V=\sum_\alpha A_\alpha\otimes B_\alpha$, with each $A_\alpha$ and
$B_\alpha$ Hermitian. The reservoir state
$\rho_B$ is stationary and centered, so that
$\langle B_\alpha\rangle_B=0$. Write
$C_{\alpha\beta}(u)=\langle B_\alpha(u)B_\beta\rangle_B$.
We assume these correlations are locally integrable and that the
second-order reduced expression below is well defined. This formulation
also allows a stationary reservoir state specified by its expectations.
With the coupling constant excluded from $V$, the full second cumulant is
\begin{equation}\label{eq:microK2}
 K_2(t)X=-\int_0^t ds\int_0^s du\,
 \tr_B[V(s),[V(u),X\otimes\rho_B]].
\end{equation}
It includes the Hamiltonian contribution, as in the ordinary second-order
construction of Ref.~\cite{Rivas2017}. The self-consistent renormalized
construction of Ref.~\cite{WinczewskiAlicki2021} uses a different
reference Hamiltonian and compensating counterterms. Its secular slope
and remainder must be identified in that same convention; deleting the
Hamiltonian part of \eqref{eq:microK2} alone is not a change of picture.
For each fixed $t$, the ordinary cumulant's dissipative
coefficient matrix is a positive Gram matrix of integrated reservoir
operators; hence $e^{\lambda^2K_2(t)}$ is CPTP
\cite{SchallerBrandes2008,Rivas2017}. This property does not imply
CP divisibility of the family.

\begin{proposition}[Bounded cumulant remainder]\label{prop:cumulant}
Suppose $C_{\alpha\beta}\in L^1(0,\infty)$ for every pair.
For the fixed system Hamiltonian, there is a secular Davies generator
$L_{\rm D}$ such that
$K_2(t)=tL_{\rm D}+R_t$, with $\norm{R_t}=o(t)$.
If also
\begin{equation}\label{eq:correlationmoment}
 \int_0^\infty u\,|C_{\alpha\beta}(u)|\,du<\infty
 \quad\text{for every }\alpha,\beta,
\end{equation}
then $\sup_{t\ge0}\norm{R_t}<\infty$.
The remainder is Hermiticity-preserving and trace-annihilating.
If $L_{\rm D}$ is relaxing, $e^{\lambda^2K_2(t)}$ converges to its
replacement channel for every fixed $\lambda\ne0$; under
\eqref{eq:correlationmoment} its state error is $O(t^{-1})$ uniformly
over input states, with the leading term in \eqref{eq:cumulantresponse}.
\end{proposition}

\begin{proof}
Decompose $A_\alpha(s)=\sum_\omega e^{-i\omega s}A_\alpha(\omega)$.
After setting $v=s-u$ in \eqref{eq:microK2}, collect equal frequencies
$\Delta=-(\omega+\nu)$. Equivalently, these are differences of Bohr
frequencies in the usual adjoint-operator convention. Define
\begin{equation}\label{eq:microF}
\begin{split}
 F_\Delta(v)X=-\!\sum_{\substack{\alpha,\beta,\omega,\nu\\
                         \omega+\nu=-\Delta}}
 e^{i\nu v}\bigl\{
 &C_{\alpha\beta}(v)[A_\alpha(\omega),A_\beta(\nu)X]\\
 {}+{}&C_{\beta\alpha}(-v)[XA_\beta(\nu),A_\alpha(\omega)]
 \bigr\}.
\end{split}
\end{equation}
There are finitely many terms, and
$C_{\beta\alpha}(-v)=C_{\alpha\beta}(v)^*$.
Thus each $F_\Delta$ is integrable in induced trace norm, and the same
is true of $vF_\Delta(v)$ under \eqref{eq:correlationmoment}.
Integration over the finite triangle gives
\begin{equation}\label{eq:microtriangle}
 K_2(t)=\int_0^t(t-v)F_0(v)\,dv+
 \sum_{\Delta\ne0}\int_0^t
 \frac{e^{i\Delta t}-e^{i\Delta v}}{i\Delta}F_\Delta(v)\,dv.
\end{equation}
The secular operator is $L_{\rm D}=\int_0^\infty F_0(v)\,dv$.
It retains all equal-frequency terms, including degeneracies and the
secular Hamiltonian correction. This is the usual weak-coupling
generator\cite{Davies1974}; positivity of the reservoir spectral
matrices gives its GKSL form. In particular,
\begin{equation}\label{eq:microremainderbound}
\begin{split}
 \norm{R_t}\le{}&\int_0^\infty\min(t,v)\norm{F_0(v)}\,dv\\
 &+\sum_{\Delta\ne0}\frac{2}{|\Delta|}
                \int_0^\infty\norm{F_\Delta(v)}\,dv.
\end{split}
\end{equation}
Dividing by $t$ and applying dominated convergence proves $R_t=o(t)$.
The first-moment assumption makes the same bound uniform in $t$.
The structural properties follow by subtraction from $K_2(t)$.
Apply Theorem~\ref{thm:main} and Corollary~\ref{cor:response} with
$L=\lambda^2L_{\rm D}$ and $D_t=\lambda^2R_t$ to obtain the relaxation
statements. The factors $\lambda^2$ cancel in the leading displacement.
\end{proof}

The hypothesis is sufficient, not necessary: an oscillatory first moment
of $F_0$ may converge even when its absolute first moment does not.
Conversely, integrability alone cannot ensure boundedness. For example,
the positive-definite classical correlation $C(v)=(1+|v|)^{-2}$ and a
commuting Hermitian system coupling $A$ give
$K_2(t)=-[t-\log(1+t)][A,[A,\cdot]]$.
Here $R_t=\log(1+t)[A,[A,\cdot]]$ is unbounded. This dephasing example
illustrates the correlation criterion; its secular generator is not
relaxing to a unique state.

Even with a finite first moment, $R_t$ need not have a limit. In fact,
put $B_\Delta=\int_0^\infty F_\Delta(v)dv$ and
$J_\Delta=\int_0^\infty e^{i\Delta v}F_\Delta(v)dv$.
Equation~\eqref{eq:microtriangle} gives
\begin{equation}\label{eq:microoscillatory}
 R_t=-\int_0^\infty vF_0(v)dv+
 \sum_{\Delta\ne0}\frac{e^{i\Delta t}B_\Delta-J_\Delta}{i\Delta}
 +o(1).
\end{equation}
The nonzero-frequency terms can persist. The constants in
\eqref{eq:microremainderbound} concern a fixed spectrum; they need not
be uniform when distinct Bohr frequencies approach one another.
No conclusion for unbounded oscillator couplings follows from this
finite-dimensional argument without additional domain and norm estimates.

\subsection{Inverse-time relaxation of the transverse qubit}\label{app:rivasqubit}
Let $H_0=\Omega\sigma_z/2$ and $V=\sigma_x\otimes B$, and fix $\lambda\ne0$.
The integrated excitation and emission coefficients of $K_2(t)$ are
$k_+(t)$ and $k_-(t)$, respectively. Put $S_t=k_+(t)+k_-(t)$.
Populations and coherences evolve in separate blocks. Exponentiating
the population block gives the excited population for any input with
initial excited population $p_0$~\cite{Rivas2017,SuarezEtAl2024},
\begin{equation}\label{eq:cumulantpopulation}
 p_t=\frac{k_+(t)}{S_t}(1-e^{-\lambda^2S_t})+p_0 e^{-\lambda^2S_t}
 \qquad (S_t>0).
\end{equation}
In the vacuum write $C(s)=\int_0^\infty J(\nu)e^{-i\nu s}d\nu$,
with the overall coupling $\lambda^2$ kept outside $K_2$. The spectral formulas
of Ref.~\cite[Eqs.~(10) and (11)]{Rivas2017} then read
\begin{equation}\label{eq:rivasspectral}
 k_\pm(t)=2\int_0^\infty J(\nu)
 \frac{1-\cos[(\nu\pm\Omega)t]}{(\nu\pm\Omega)^2}\,d\nu .
\end{equation}
For the Ohmic $J$ of Sec.~\ref{sec:cumulant}, $J(\nu)/(\nu+\Omega)^2$
is integrable. The Riemann--Lebesgue lemma gives
\begin{equation}\label{eq:rivaslimits}
 k_+(t)\longrightarrow d_+=2\int_0^\infty
 \frac{J(\nu)}{(\nu+\Omega)^2}\,d\nu>0 .
\end{equation}
For emission, $2(1-\cos[xt])/(t x^2)$ is an approximate identity
of mass $2\pi$ on the real line. Extending $J$ by zero to negative
frequencies and using continuity at $\Omega$ yields
$k_-(t)/t\to g_-=2\pi J(\Omega)>0$. Equation~\eqref{eq:cumulantpopulation}
now gives $t p_t\to d_+/g_-$ uniformly in $p_0$. The Ohmic correlation
$C(s)=\alpha(\omega_c^{-1}+is)^{-2}$ has a divergent first absolute
moment, so the population calculation goes beyond \eqref{eq:correlationmoment}.

To pass from populations to trace norm, write $z_t=(\widetilde\Phi_t
\rho)_{eg}$. The Lamb shift is diagonal. Along the auxiliary parameter
$0\le u\le1$ of $e^{u \lambda^2 K_2(t)}$, the coherence $z$ therefore obeys
$dz/du=\lambda^2[(-S_t/2+ih_t)z+m_t\bar z]$ for real $h_t$.
Positivity of the integrated coefficient matrix gives
$|m_t|\le\sqrt{k_+(t)k_-(t)}$~\cite{Rivas2017}. Taking the real part
of $\bar z\,dz/du$ and integrating bounds the output by
\begin{equation}\label{eq:qubitcohbound}
 |z_t|\le |z_0|\exp\!\left[-\tfrac{\lambda^2}{2}
  \big(\sqrt{k_-(t)}-\sqrt{k_+(t)}\big)^2\right].
\end{equation}
The exponent is $-\lambda^2g_-t/2+o(t)$ and $|z_0|\le1/2$. Since the trace
norm to $\pi_g$ is $2\sqrt{p_t^2+|z_t|^2}$, its coefficient is
$2d_+/g_-$ for every input. To evaluate it, put $\chi=\Omega/\omega_c$
and $E_1(\chi)=\int_\chi^\infty e^{-v}dv/v$. Substitution of
$u=\nu/\Omega$ in \eqref{eq:rivaslimits} gives
\begin{align*}
 \Omega\,\frac{2d_+}{g_-}
 &=\frac{2e^\chi}{\pi}\int_0^\infty\frac{u e^{-\chi u}}{(1+u)^2}\,du\\
 &=\frac{2e^\chi}{\pi}\big[(1+\chi)e^\chi E_1(\chi)-1\big].
\end{align*}
At $\chi=1/5$ this is \eqref{eq:rivasqubittail}.

The same spectral integral determines the leading shift of the
interacting ground-state population. The first-order ground-state
correction has a bare excited qubit and one boson, with energy
denominator $\Omega+\nu$. Its squared norm gives
\[
 p_e^{\rm gs}(\lambda)
 =\lambda^2\int_0^\infty\frac{J(\nu)}{(\Omega+\nu)^2}\,d\nu
  +O(\lambda^4).
\]
Existence and analyticity of the ground state at small coupling follow
from Ref.~\cite[Theorem~1]{HaslerHerbst2011}; Corollary~5 there obtains
the expansion coefficients as limits of infrared-regularized
perturbation theory.
For a radial realization of this Ohmic spectrum, the form factor
$f$ in that reference is proportional to $e^{-\nu/(2\omega_c)}$.
Evenness in the coupling follows by conjugating with the field parity operator,
which sends $\lambda$ to $-\lambda$ and leaves the population
observable invariant. The leading term equals
$\lambda^2d_+/2\simeq0.03960$ at the parameters of
Fig.~\ref{fig:rivasqubit}. The small-coupling relaxation theorem of
Ref.~\cite{DeRoeckKupiainen2013} applies because
$C(s)=O(s^{-2})$ has a finite absolute moment of every order in
$(0,1)$, and $J(\Omega)>0$ gives the Fermi golden rule condition.
Theorem~1.1 there yields the long-time limit for every initial vector
and every system observable at sufficiently small fixed coupling. Since
the ground state exists, Remark~1.2 there identifies this limit with
the ground-state expectation; the additional regularity conditions on
the initial state and observable are needed only for the rate of
convergence. The value
$0.03960$ evaluates the second-order term at the plotting parameters.

\subsection{A finite-temperature check with a Drude bath}\label{app:drudequbit}
For a thermal correlation $C(s)=\langle B(s)B\rangle_B$ the coefficients
are $k_\pm(t)=2\Re\int_0^t(t-s)e^{\mp i\Omega s}C(s)\,ds$.
Under \eqref{eq:correlationmoment} this gives
$k_\pm(t)=t g_\pm+d_\pm+o(1)$, where
\begin{equation}\label{eq:cumulantpopulationconstants}
\begin{split}
 g_\pm&=2\Re\int_0^\infty e^{\mp i\Omega s}C(s)\,ds,\\
 d_\pm&=-2\Re\int_0^\infty s e^{\mp i\Omega s}C(s)\,ds .
\end{split}
\end{equation}
Writing $g=g_++g_->0$ and $p_*=g_+/g$, expansion of
\eqref{eq:cumulantpopulation} yields
\begin{equation}\label{eq:cumulantpopulationrate}
 t(p_t-p_*)\longrightarrow
 C_{\rm pop}=\frac{(1-p_*)d_+-p_*d_-}{g}.
\end{equation}
Choose units $\Omega=1$, inverse temperature $\beta=5$ and Drude cutoff
$\gamma_{\rm D}=2$. With $J(\nu)=2\gamma_{\rm D}\nu/[\pi(\nu^2+\gamma_{\rm D}^2)]$, the
Matsubara expansion is $C(s)=\sum_{n\ge0}c_n e^{-\nu_n s}$ for $s>0$.
The common bath strength is absorbed into $\lambda^2$. In this convention the
frequencies and weights are~\cite[Eqs.~(31)--(35)]{YinEtAl2012}
\begin{align*}
 \nu_0&=\gamma_{\rm D}, & c_0&=\gamma_{\rm D}[\cot(\beta\gamma_{\rm D}/2)-i],\\
 \nu_n&=2\pi n/\beta, & c_n&=\frac{4\gamma_{\rm D}\nu_n}{\beta(\nu_n^2-\gamma_{\rm D}^2)}
 \quad(n\ge1).
\end{align*}
Detailed balance and the spectral density at resonance give
$p_*=1/(1+e^\beta)$ and $g=4\gamma_{\rm D}\coth(\beta/2)/(1+\gamma_{\rm D}^2)$.
Writing $z_{n,\pm}=\nu_n\pm i$, integration gives
\[
 k_\pm(t)=t g_\pm-2\Re\sum_{n\ge0}c_n
 \frac{1-e^{-t z_{n,\pm}}}{z_{n,\pm}^2}.
\]
Thus $d_\pm=-2\Re\sum_{n\ge0}c_n/z_{n,\pm}^2$.
The sums are absolutely convergent. For $\nu_n\ge2\gamma_{\rm D}$,
$|c_n|\le16\gamma_{\rm D}/(3\beta\nu_n)$. Inserting this bound in
\eqref{eq:cumulantpopulationrate} and using
$\sum_{n>N}n^{-3}\le1/(2N^2)$ yields, once $\nu_{N+1}\ge2\gamma_{\rm D}$,
\[
 |C_{\rm pop}-C_{\rm pop}^{(N)}|
 \le\frac{16\gamma_{\rm D}}{3\beta g N^2}
       \left(\frac{\beta}{2\pi}\right)^3 .
\]
Keeping $n\le10^5$ gives $C_{\rm pop}^{(N)}=0.277658673066$ with
truncation error below $6.7\times10^{-11}$. Here $g_->g_+$, so
\eqref{eq:qubitcohbound} again suppresses coherences exponentially.
The limit of $t\norm{\widetilde\Phi_t\rho-\rho_\beta}_1$ is therefore
$2|C_{\rm pop}|\simeq0.55531735$ for every
input, where $\rho_\beta$ is the bare Gibbs state. As a finite-time
check, evaluating \eqref{eq:cumulantpopulation} at $\lambda=0.2$ gives
$t(p_t-p_*)=0.2777245$, $0.2776806$ and $0.2776653$ at
$t=1000$, $3000$ and $10000$, respectively, for both the ground and
the excited preparation to the precision shown.

\subsection{Why the Redfield propagator is different}
For the same second-order expansion, the finite-upper-limit,
interaction-picture Redfield equation is
$\dot\rho=\lambda^2\dot K_2(t)\rho$. Its propagator is
$\mathcal T\exp[\lambda^2\int_0^t\dot K_2(s)ds]$, whereas the
cumulant approximation uses the ordinary exponential
$e^{\lambda^2K_2(t)}$. Differentiation gives the exact identity
\cite{Rivas2017,WinczewskiAlicki2021}
\begin{equation}\label{eq:cumulantgenerator}
 \mathcal G_{\rm cum}(t)=\lambda^2\int_0^1
 e^{u\lambda^2K_2(t)}\dot K_2(t)e^{-u\lambda^2K_2(t)}du.
\end{equation}
At fixed time this is
$\lambda^2\dot K_2+\lambda^4[K_2,\dot K_2]/2+O(\lambda^6)$.
The agreement to second order is therefore a fixed-time perturbative
statement, not an all-time equivalence of the two propagators.
Commutation of the instantaneous generators is a sufficient special
case in which time ordering can be removed.

For a time-independent nonsecular Bloch--Redfield generator
$\mathcal L_{\rm BR}$ in the Schr\"odinger picture, writing
$e^{t\mathcal L_{\rm BR}}=e^{tL+t(\mathcal L_{\rm BR}-L)}$
gives a correction linear in $t$ at fixed coupling. Thus our
sublinear-perturbation theorem does not identify its limiting state
with that of $L$. If its restriction to the trace-zero subspace is
Hurwitz and it has a normalized stationary vector, its own
finite-dimensional semigroup converges exponentially to that vector;
positivity requires a separate check. A fully secular Davies equation
instead fits our setting with $D_t=0$ whenever it is relaxing.
Other coarse-grained or completely positive variants must be checked
against their actual generator or exponent.

% APS bibliography generated from the separate BibTeX database.
% REVTeX selects apsrev4-2 and supplies the longbibliography controls.
\bibliography{references}
\end{document}